\documentclass{article} 
\usepackage{iclr2026_conference,times}

\usepackage{amsmath,amsfonts,bm}

\def\eqref#1{equation~\ref{#1}}

\def\1{\bm{1}}

\DeclareMathAlphabet{\mathsfit}{\encodingdefault}{\sfdefault}{m}{sl}
\SetMathAlphabet{\mathsfit}{bold}{\encodingdefault}{\sfdefault}{bx}{n}

\usepackage{hyperref}
\usepackage{url}
\usepackage{graphicx}
\usepackage{multicol}
\usepackage{multirow}
\usepackage{booktabs}
\usepackage{makecell}

\usepackage{fancyhdr}

\usepackage{textcomp, gensymb, amssymb}
\usepackage[acronym]{glossaries}
\usepackage{algorithm, algpseudocode}
\usepackage{subcaption}
\usepackage{footnote}
\usepackage{pdfpages}
\providecommand{\tightlist}{%
  \setlength{\itemsep}{0pt}\setlength{\parskip}{0pt}}
\usepackage{fancyvrb}

\title{Learning-guided Kansa collocation for forward and inverse PDEs beyond linearity}

\author{Zheyuan Hu$^{1}$, Weitao Chen$^{2}$, Cengiz Öztireli$^{1}$, Chenliang Zhou$^{1*}$, Fangcheng Zhong$^{1*}$\\
  $^{1}$ Department of Computer Science and Technology, \\
  $^{2}$ Department of Applied Mathematics and Theoretical Physics,\\
  University of Cambridge, UK. \quad $^{*}$Co-corresponding authors.\\
  {\tt \small  \{zh369, wc358\}@cam.ac.uk, \{chenliang.zhou, fangcheng.zhong\}@cst.cam.ac.uk
  }
}

\iclrfinalcopy 

\newacronym{brdf}{BRDF}{Bidirectional Reflectance Distribution Function}
\newacronym{bsdf}{BSDF}{Bidirectional Scattering Distribution Function}
\newacronym{pde}{PDE}{Partial Differential Equation}
\newacronym{fdm}{FDM}{Finite Difference Method}
\newacronym{fem}{FEM}{Finite Element Method}
\newacronym{pinn}{PINN}{Physics-Informed Neural Network}
\newacronym{fno}{FNO}{Fourier Neural Operator}
\newacronym{mlp}{MLP}{Multi-Layer Perceptron}
\newacronym{uml}{UML}{Unified Modeling Language}
\newacronym{rbf}{RBF}{Radial Basis Function}

\begin{document}
\maketitle
\begin{abstract}
Partial Differential Equations are precise in modelling the physical, biological and graphical phenomena. However, the numerical methods suffer from the curse of dimensionality, high computation costs and domain-specific discretization. We aim to explore pros and cons of different PDE solvers, and apply them to specific scientific simulation problems, including forwarding solution, inverse problems and equations discovery. In particular, we extend the recent \cite{zhong2023CNF} framework solver to coupled and non-linear settings, together with down-stream applications. The outcomes include implementation of selected methods, self-tuning techniques, evaluation on benchmark problems and a comprehensive survey of neural PDE solvers and scientific simulation applications.
\end{abstract}

\section{Introduction}

PDEs are useful in different domains of scientific computing, including physics, graphics and biology.
\cite{zhong2023CNF} proposed extension to Kansa method, which is a mesh-free Radial Basis Functions (RBFs) PDE solver. They introduced auto-tuning of the shape parameters of RBFs. However, their work focuses only on \textbf{single-variable linear PDEs}. Therefore, this paper extends CNFs backend solver to \textbf{multiple unknown functions \(u\) and nonlinear PDEs}, and apply the framework to specific scientific simulation problems, including forward computation and inverse problems.


It's unknown how (extended) Constrained Neural Fields (CNF) \cite{zhong2023CNF} compared with other classical and neural PDE solvers. Hence, we also implement and evaluate selected prior methods on the benchmarks on their \textbf{effectiveness} with different
quality metrics (e.g.~L1, L2, errors) against ground truth solutions, \textbf{efficiency}, computation resource, convergence speed, \textbf{method complexity}, and finally \textbf{utility in research},
i.e.~their scientific simulation applications or integration with other methods, e.g.~differentiable rendering to solve inverse physics-related problems in Graphics.


\section{Related work}

\textbf{PDE benchmarks.} We identified several representative equations \cite{Takamoto2022PDEBench} in Table \ref{tab:pde_summary}. They are different in linearity of the operator and solution \(u\) dimensionality.


\textbf{PDE solvers}. Numerical methods, e.g.~Finite Difference Method (FDM) and Finite Element Method (FEM), are widely used to solve PDEs. However, they suffer from the curse of dimensionality, high computation costs and domain-specific discretization. Recently, neural network based solvers have shown promising results in addressing these issues. For example, Physics-Informed Neural Networks (PINNs) \cite{RAISSI2019PINNs} and Fourier Neural Operators (FNOs) \cite{Li2020FNO} have demonstrated the ability to generalize to unseen scenarios and handle high dimension effectively.

\textbf{Inverse problem}, i.e.~estimating unknown parameters or inputs of a variable \(x\) from given solution observations \(u\), is crucial. However, it's unclear how CNFs can be applied to these problems, including connecting with differentiable rendering pipelines \cite{spielberg2023DiffVC} in Visual Computing.


\section{Methodology}

\hypertarget{general-form-of-pdes}{%
\subsection{General form of PDEs}\label{general-form-of-pdes}}

With spatial domain $\Omega \subset \mathbb{R}^d$, where its dimension is $d$, and the unknown field
\(u(x, t) \in \mathcal{U}: \mathbb{D} \to \mathbb{R}\) defined on the
spatio-temporal
domain
\(\mathbb{D} = \Omega \times [t_0, t_f] \subset \mathbb{R}^{d+1}\), the
general form of PDEs is,
\begin{equation}
\begin{cases}
\begin{aligned}
\mathcal{D}[u] = f, & \quad x \in \Omega, t \in [t_0, t_f], \\
\mathcal{B}_i[u] = g_i, & \quad x \in \partial \Omega_i, t \in  [t_0, t_f].
\end{aligned} 
\end{cases} 
\Leftrightarrow
\begin{cases}
\begin{aligned}
\mathcal{D}[u](x, t) = f(x, t),  & \quad x \in \Omega, t \in [t_0, t_f], \\
\mathcal{B}_i[u](x, t) = g_i(x, t) & \quad x \in \partial \Omega_i, t \in  [t_0, t_f].
\end{aligned} .
\end{cases} \\
\label{eq:pde-general}
\end{equation}
where \(\mathcal{D}: \mathcal{U} \to \mathcal{Y}\) is the differential
operator and \(f \in \mathcal{Y}: \mathbb{D} \to \mathbb{R}^m\) is source function, e.g.~the external force in dynamics, with \(m\)
being the output dimension of
\(f\)\footnote{Note that $\mathcal{U}$ and $\mathcal{Y}$ are two function spaces, and we require they are Banach spaces.}.
The differential operators \(\mathcal{D}\) include the gradient
\(\nabla\), Laplace \(\Delta\), divergence \(\nabla \cdot\), etc. For
boundary conditions, \(\mathcal{B}_i: \mathcal{U} \to \mathcal{Z}_i\) is
each boundary operator with
\(g_i \in \mathcal{Z}_i: \partial \Omega_i \times [t_0, t_f] \to \mathbb{R}^{n_i}\)
and \(n_i\) as the output dimension of \(g_i\).



\hypertarget{kansa-collocation}{%
\subsection{Kansa collocation}\label{kansa-collocation}}

\textbf{Kernel functions}. \acrfull{rbf} relates the distance \(r\) between the input \(\mathbf{x}\) and a fixed origin point
\(\mathbf{c}\) to the output value.
\begin{equation}
\psi_{\mathbf{c}}( r ) = \psi_{\mathbf{c}}(|| \mathbf{x} - \mathbf{c} ||).
\label{eq:rbf-general}
\end{equation}
There are various infinitely smooth RBFs, among which we choose Gaussian
RBF for its effectiveness in approximating smooth functions,
\begin{equation}
\psi_{\mathbf{c}}(r) = 
\begin{cases}
e^{-(\epsilon r)^2}, & \quad \text{Gaussian},
\\
\frac{1}{1 + (\epsilon r)^2}, & \quad \text{Inverse quadratic},
\\
\sqrt{1 + (\epsilon r)^2}, & \quad \text{Multiquadrics},
\end{cases}
\label{eq:rbf-types}
\end{equation}
where Gaussian shape parameter \(\epsilon = \frac{1}{\sqrt{2} \sigma}\),
and \(\sigma\) is the standard deviation.

\textbf{Kansa method} \cite{kansa1990multiquadrics2} approximates the
solution \(u(x, t)\) with a linear combination of kernel functions
\(\psi_k(|| \mathbf{x}_i - \mathbf{x}_k ||) \in \mathbb{R}^d \to \mathbb{R}\)
centered at each collocation point
\(\{\mathbf{x}_i \in \mathbb{D} \}_{i=1}^N\),
\begin{equation}
u(x_i, t_i) \approx \hat{u}(\mathbf{x}_i) = \sum_{k=1}^{N} \alpha_k \cdot \psi_k(|| \mathbf{x}_i - \mathbf{x}_k ||), \quad \mathbf{x} \in \mathbb{D},
\label{eq:kansa-approximation}
\end{equation}
where \(\alpha_i \in \mathbb{R}\) are the coefficients to be solved. The time dimension $t$ is omitted, which can be treated as an additional spatial dimension here. Equation \eqref{eq:kansa-approximation} is expressed as, by rewriting
the kernel functions into matrix form,
\begin{equation}
\underbrace{
\begin{bmatrix}  
\hat{u}(\mathbf{x}_1) \\
\hat{u}(\mathbf{x}_2) \\
\vdots \\
\hat{u}(\mathbf{x}_N)
\end{bmatrix}
}_{\mathbf{u} \in \mathbb{R}^{N}}
=
\underbrace{
\begin{bmatrix}
\psi_1(|| \mathbf{x}_1 - \mathbf{x}_1 ||) & \psi_2(|| \mathbf{x}_1 - \mathbf{x}_2 ||) & \cdots & \psi_N(|| \mathbf{x}_1 - \mathbf{x}_N ||) \\
\psi_1(|| \mathbf{x}_2 - \mathbf{x}_1 ||) & \psi_2(|| \mathbf{x}_2 - \mathbf{x}_2 ||) & \cdots & \psi_N(|| \mathbf{x}_2 - \mathbf{x}_N ||) \\
\vdots & \vdots & \ddots & \vdots \\
\psi_1(|| \mathbf{x}_N - \mathbf{x}_1 ||) & \psi_2(|| \mathbf{x}_N - \mathbf{x}_2 ||) & \cdots & \psi_N(|| \mathbf{x}_N - \mathbf{x}_N ||)
\end{bmatrix}
}_{\text{kernel matrix } \mathbf{K} \in \mathbb{R}^{N \times N}}
\cdot
\underbrace{
\begin{bmatrix}
\alpha_1 \\
\alpha_2 \\
\vdots \\
\alpha_N
\end{bmatrix}
}_{\mathbf{a} \in \mathbb{R}^{N}}.
\label{eq:kansa-matrix}
\end{equation}
The PDE general form \eqref{eq:pde-general} can be summarized as a
single equation,
\begin{equation}
\mathcal{F}[\hat{u}](\mathbf{x}_i) = h(\mathbf{x}_i), \quad \mathbf{x}_i \in \mathbb{D},
\label{eq:pde-kansa}
\end{equation}
where the operator \(\mathcal{F} = \{\mathcal{D}, \mathcal{B}_i\}\) and \(h = \{f, g_i\}\) represent both the initial
and boundary conditions.

\hypertarget{linear-operator-case}{%
\subsubsection{Linear operator case}\label{linear-operator-case}}

By plugging in the approximation of \(u\) \eqref{eq:kansa-approximation}
and assuming the operator \(\mathcal{F}\) is
\emph{linear}\footnote{For linear operators, $\mathcal{F}[\alpha \cdot \psi] = \alpha \cdot \mathcal{F}[\psi]$, as defined in § \ref{linear-operator}.},
the PDE \eqref{eq:pde-kansa} can be simplified as,
\begin{equation}
\mathcal{F}[\hat{u}](\mathbf{x}_i) = \mathcal{F}[\sum_{k=1}^{N} \alpha_k \cdot \psi_k](\mathbf{x}_i) = \sum_{k=1}^{N} \alpha_k  \cdot \mathcal{F}[\psi_k](\mathbf{x}_i) = h(\mathbf{x}_i).
\label{eq:pde-kansa-linear}
\end{equation}
By expanding in matrix form, the above equation is,
\begin{equation}
\underbrace{
\begin{bmatrix}
\mathcal{F}[\psi_1](\mathbf{x}_1) & \mathcal{F}[\psi_2](\mathbf{x}_1) & \cdots & \mathcal{F}[\psi_N](\mathbf{x}_1) \\
\mathcal{F}[\psi_1](\mathbf{x}_2) & \mathcal{F}[\psi_2](\mathbf{x}_2) & \cdots & \mathcal{F}[\psi_N](\mathbf{x}_2) \\   
\vdots & \vdots & \ddots & \vdots \\
\mathcal{F}[\psi_1](\mathbf{x}_N) & \mathcal{F}[\psi_2](\mathbf{x}_N) & \cdots & \mathcal{F}[\psi_N](\mathbf{x}_N)
\end{bmatrix}
}_{\text{operator-evaluated kernel matrix }\mathbf{F} \in \mathbb{R}^{N \times N}}
\cdot
\underbrace{
\begin{bmatrix}
\alpha_1 \\
\alpha_2 \\
\vdots \\
\alpha_N
\end{bmatrix}
}_{\mathbf{a} \in \mathbb{R}^{N}}
=
\underbrace{
\begin{bmatrix}
h(\mathbf{x}_1) \\
h(\mathbf{x}_2) \\
\vdots \\
h(\mathbf{x}_N)
\end{bmatrix}
}_{\text{constraint values }\mathbf{h} \in \mathbb{R}^{N}}.
\label{eq:kansa-pde-matrix}
\end{equation}
Concretely, when the kernel function is Gaussian RBF defined in
\eqref{eq:rbf-general}, and \(\mathbf{x}_i = (x_i, t_i)\),
\begin{equation}
\psi_k = e^{-\frac{r_k^2}{2 \sigma^2}}, \quad  r_k^2 = (x_k-x_i)^2 + (t_k-t_i)^2.
\end{equation}
Take \(\mathcal{F} = \frac{\partial }{\partial t}\)
\footnote{$\frac{\partial \psi_k}{\partial t}$: \texttt{phi\_t = torch.autograd.grad(phi, t, create\_graph=True)[0]}}
and by the chain rule, the element \(\mathbf{F}_{k, i}\) in the matrix
\eqref{eq:kansa-pde-matrix} is thus,
\begin{equation}
\mathcal{F}[\psi_k](\mathbf{x}_i)  = \frac{\partial \psi_k (\mathbf{x}_i)}{\partial t} 
= \frac{\partial \psi_k(\mathbf{x}_i)}{\partial r_k^2} \cdot \frac{\partial r_k^2}{\partial t} 
= -\frac{1}{2\sigma^2} e^{-\frac{r_k^2}{2\sigma^2}} \cdot 2(t_k - t_i)
= -\frac{t_k -  t_i}{\sigma^2} e^{-\frac{r_k^2}{2\sigma^2}}.
\end{equation}
\textbf{Simultaneous equations.} When there are \(N_{eq}\) equations of
different constraints to be satisfied, the collocation points
\(\{\mathbf{x}_i \in \mathbb{D} \}_{i=1}^{N_{\text{total}}}\) are
distributed among all
equations\footnote{Note that repeated collocation points are forced to be repeated here for distinct constraints.},
where \(N_{\text{total}} = \sum_{j=1}^{N_{eq}} N_j\).
\begin{equation}
\mathcal{F}_j[\hat{u}](\mathbf{x}_i) = h_j(\mathbf{x}_i), \quad \forall j \in \{1, \ldots, N_{eq}\},  \mathbf{x}_i \in \mathbb{D}.
\label{eq:pde-kansa-linear-multiple-eq}
\end{equation}
Equation \eqref{eq:kansa-pde-matrix} can be extended by stacking each
matrix
\(\mathbf{F}^{(j)} \in \mathbb{R}^{N_{\text{total}} \times N_{\text{total}}}\)
and constraint vector
\(\mathbf{h}^{(j)} \in \mathbb{R}^{N_{\text{total}}}\) vertically for
all equations \cite{zhong2023CNF}. The block matrix form is,
\begin{equation}
\underbrace{
\begin{bmatrix}
\mathbf{F}^{(1)} \\
\vdots \\
\mathbf{F}^{(N_{eq})}
\end{bmatrix}
}_{\text{stacked } \mathbf{F} \in \mathbb{R}^{(N_{eq} \cdot N_{\text{total}}) \times N_{\text{total}}}}
\cdot
\underbrace{
\begin{bmatrix}
\alpha_1 \\
\vdots \\
\alpha_{N_{\text{total}}}
\end{bmatrix}
}_{\mathbf{a} \in \mathbb{R}^{N_{\text{total}}}}
=
\underbrace{
\begin{bmatrix}
\mathbf{h}^{(1)} \\
\vdots \\
\mathbf{h}^{(N_{eq})}
\end{bmatrix}
}_{\text{stacked } \mathbf{h} \in \mathbb{R}^{N_{eq} \cdot N_{\text{total}}}}.
\label{eq:kansa-pde-matrix-multiple-eq}
\end{equation}
\textbf{The solution} of \(u\) \eqref{eq:kansa-approximation} depends on
the coefficients \(\mathbf{a}=[\alpha_1, \alpha_2, \ldots, \alpha_N]\),
which can be solved by the linear system
\(\mathbf{F} \mathbf{a} = \mathbf{h}\). The general form is given by the
least squares approximation, i.e.~minimizing the norm of the error
vector and setting the gradient to zero,
\begin{equation}
\begin{aligned}
\mathbf{a}^{\text{opt}} 
&= \min_{\mathbf{a}} (||\mathbf{F} \mathbf{a} - \mathbf{h}||)^2, \\
& \nabla_{\mathbf{a}}(\mathbf{F} \mathbf{a} - \mathbf{h})^T (\mathbf{F} \mathbf{a} - \mathbf{h}) = 0 \implies  
(\mathbf{F}^T \mathbf{F})\mathbf{a}^{\text{opt}}  = \mathbf{F}^T \mathbf{h}.
\label{eq:kansa-least-squares-solution}
\end{aligned}
\end{equation}
If matrix $\mathbf{F}$ is full rank, $\mathbf{F}^T \mathbf{F}$ is invertible, thus one can derive $\mathbf{a}^{\text{opt}} = (\mathbf{F}^T \mathbf{F})^{-1} \mathbf{F}^T \mathbf{h}$. Should the matrix $\mathbf{F}$ be square and invertible, \eqref{eq:kansa-least-squares-solution} can be further simplified as $\mathbf{a}^{\text{opt}} = \mathbf{F}^{-1} \mathbf{h}$.  Whichever conditions occurs, the final solution for $u$ is approximated by plugging in the optimal coefficients $\mathbf{a}^{\text{opt}}$ into \eqref{eq:kansa-matrix}, i.e. $\hat{u}(\mathbf{x}) = \mathbf{K} \cdot \mathbf{a}^{\text{opt}}$.

When testing on unseen data points
\(\{ \mathbf{x}_j^{\star} \in \mathbb{D} \}_{j=1}^M\), the kernel
functions are constructed between the test points and the collocation
points \(\{\mathbf{x}_i \in \mathbb{D} \}_{i=1}^N\). The solution is thus,
\begin{equation}
u(x, t) \approx \hat{u}(\mathbf{x}^{\star}) = \sum_{k=1}^{N} \alpha_k \cdot \psi_k(|| \mathbf{x}^{\star} - \mathbf{x}_k ||), \quad \mathbf{x} \in \mathbb{D},
\label{eq:kansa-approximation-test}
\end{equation}
Test-time solution \eqref{eq:kansa-approximation-test} can be formulated
to matrix form,
\begin{equation}
\underbrace{
\begin{bmatrix}  
\hat{u}(\mathbf{x}^{\star}_1) \\
\hat{u}(\mathbf{x}^{\star}_2) \\
\vdots \\
\hat{u}(\mathbf{x}^{\star}_M)
\end{bmatrix}
}_{\mathbf{u} \in \mathbb{R}^{M}}
=
\underbrace{
\begin{bmatrix}
\psi_1(|| \mathbf{x}^{\star}_1 - \mathbf{x}_1 ||) & \psi_2(|| \mathbf{x}^{\star}_1 - \mathbf{x}_2 ||) & \cdots & \psi_N(|| \mathbf{x}^{\star}_1 - \mathbf{x}_N ||) \\
\psi_1(|| \mathbf{x}^{\star}_2 - \mathbf{x}_1 ||) & \psi_2(|| \mathbf{x}^{\star}_2 - \mathbf{x}_2 ||) & \cdots & \psi_N(|| \mathbf{x}^{\star}_2 - \mathbf{x}_N ||) \\
\vdots & \vdots & \ddots & \vdots \\
\psi_1(|| \mathbf{x}^{\star}_M - \mathbf{x}_1 ||) & \psi_2(|| \mathbf{x}^{\star}_M - \mathbf{x}_2 ||) & \cdots & \psi_N(|| \mathbf{x}^{\star}_M - \mathbf{x}_N ||)
\end{bmatrix}
}_{\text{kernel matrix } \mathbf{K}^{\star} \in \mathbb{R}^{M \times N}}
\cdot
\underbrace{
\begin{bmatrix}
\alpha_1 \\
\alpha_2 \\
\vdots \\
\alpha_N
\end{bmatrix}
}_{\mathbf{a} \in \mathbb{R}^{N}}.
\label{eq:kansa-matrix-test}
\end{equation}
\hypertarget{extension-1-coupled-solution-fields-of-pdes}{%
\subsubsection{Extension 1: coupled solution fields of
PDEs}\label{extension-1-coupled-solution-fields-of-pdes}}

\textbf{Coupled multi-dimensional PDE solution fields.} Assuming there
are \(N_{D}\) solution dimensions,
i.e.~\(\mathbf{u} = [u_1, u_2, \ldots, u_{N_D}]\), the Kansa approximation \eqref{eq:kansa-approximation} for each dimension is,
\begin{equation}
\hat{u}_d(\mathbf{x}) = \sum_{k=1}^{N} \alpha_{k}^{(d)} \cdot \psi_k^{(d)}(|| \mathbf{x} - \mathbf{x}_k ||), \quad \forall d \in \{1, \ldots, N_D\}.
\label{eq:kansa-approximation-multi-d}
\end{equation}
The coupled PDE equation is thus formulated as applying the coupling, or
governing operator \(\mathcal{G}\) on all dimensions of solution, which
each has its own operator \(\mathcal{F}^{(d)}\),
\begin{equation}
\mathcal{G} \left( \mathcal{F}^{(1)}[\hat{u}_1], \ldots, \mathcal{F}^{(N_D)}[\hat{u}_{N_D}] \right) (\mathbf{x}_i) = h(\mathbf{x}_i), \quad \mathbf{x}_i \in \mathbb{D}.
\label{eq:pde-kansa-multi-d}
\end{equation}
Here we assume the coupling operator \(\mathcal{G}\) is linear with each
dimension of solution,
\begin{equation}
\mathcal{G} \left( \hat{v}_1, \ldots, \hat{v}_{N_D} \right)(\mathbf{x}) = \sum_{d=1}^{N_D} \beta_d \cdot \hat{v}_d(\mathbf{x}),
\end{equation}
where \(\beta_d \in \mathbb{R}\) is the per-dimension weight. Equation
\eqref{eq:kansa-pde-matrix} can be extended by stacking each matrix
\(\mathbf{F}^{(d)} \in \mathbb{R}^{N \times N}\) horizontally for all
dimensions of solution. The block matrix form is,
\begin{equation}
\underbrace{
\begin{bmatrix}
\beta_1 \textbf{I}_N & \cdots & \beta_{N_D} \textbf{I}_N
\end{bmatrix}
}_{\boldsymbol{\beta} \in \mathbb{R}^{N \times (N_D \cdot N)}}
\circ
\underbrace{
\begin{bmatrix}
\mathbf{F}^{(1)} & \cdots & \mathbf{F}^{(N_D)}
\end{bmatrix}
}_{\text{coupling } \mathbf{F} \in \mathbb{R}^{N \times (N_D \cdot N)}}
\cdot
\underbrace{
\begin{bmatrix}   
\textbf{a}^{(1)} \\
\vdots \\
\textbf{a}^{(N_D)}
\end{bmatrix}
}_{\mathbf{a} \in \mathbb{R}^{(N_D \cdot N)}}
=
\underbrace{
\begin{bmatrix}
h(\mathbf{x}_1) \\
\vdots \\
h(\mathbf{x}_N)
\end{bmatrix}
}_{\text{stacked } \mathbf{h} \in \mathbb{R}^{N}},
\label{eq:kansa-pde-matrix-multi-d}
\end{equation}
where \(\circ\) is element-wise or Hadamard product and \(\textbf{I}_N\) is the identity matrix of size \(N\). For
simultaneous coupled PDE equations, similar to
\eqref{eq:pde-kansa-linear-multiple-eq}, they are indexed by
\(j \in \{1, \ldots, N_{eq}\}\),
\begin{equation}
\mathcal{G}_j \left( \mathcal{F}_j^{(1)}[\hat{u}_1], \ldots, \mathcal{F}_j^{(N_D)}[\hat{u}_{N_D}] \right) (\mathbf{x}_i) = h_j(\mathbf{x}_i), \quad \forall j \in \{1, \ldots, N_{eq}\}, \mathbf{x}_i \in \mathbb{D}.
\label{eq:pde-kansa-multi-d-multiple-eq}
\end{equation}
With \(N_{\text{total}}\) defined as in
\eqref{eq:kansa-pde-matrix-multiple-eq}, the block matrix form is,
\begin{equation}
\underbrace{
\begin{bmatrix}
\boldsymbol{\beta}^{(1)} \\
\vdots \\
\boldsymbol{\beta}^{(N_{eq})}
\end{bmatrix}
}_{\boldsymbol{\beta} \in \mathbb{R}^{(N_{eq} \cdot N_{\text{total}}) \times (N_D \cdot N_{\text{total}})}}
\circ
\underbrace{
\begin{bmatrix}
\mathbf{F}^{(1, 1)} & \cdots & \mathbf{F}^{(1, N_D)} \\
\vdots &   \mathbf{F}^{(j, d)}  & \vdots \\
\mathbf{F}^{(N_{eq}, 1)} & \cdots & \mathbf{F}^{(N_{eq}, N_D)}
\end{bmatrix}
}_{\text{coupling } \mathbf{F} \in \mathbb{R}^{(N_{eq} \cdot N_{\text{total}}) \times (N_D \cdot N_{\text{total}})}}
\cdot
\underbrace{
\begin{bmatrix}   
\textbf{a}^{(1)} \\
\vdots \\
\textbf{a}^{(N_D)}
\end{bmatrix}
}_{\mathbf{a} \in \mathbb{R}^{(N_D \cdot N_{\text{total}})}}
=
\underbrace{
\begin{bmatrix}
\textbf{h}^{(1)} \\
\vdots \\
\textbf{h}^{(N_{eq})}
\end{bmatrix}
}_{\text{stacked } \mathbf{h} \in \mathbb{R}^{(N_{eq} \cdot N_{\text{total}})} }.
\label{eq:kansa-pde-matrix-multi-d-multiple-eq}
\end{equation}
\hypertarget{extension-2-nonlinear-operator-case}{%
\subsubsection{Extension 2: nonlinear operator
case}\label{extension-2-nonlinear-operator-case}}
When the operator \(\mathcal{F}\) is \emph{nonlinear}, we can no longer
simplify \eqref{eq:pde-kansa} as in \eqref{eq:pde-kansa-linear}.
However, we can still derive the relation between the solution \(u\) and
its linear transformed version as below.

\textbf{Differentiable matrix} helps decompose the general non-linear
operator \(\mathcal{F}\) into a series of linear operators. Take any
linear operator, e.g.~\(\frac{\partial }{\partial x}\), it relates the
relation between unknown \(\textbf{u}\) and its derivative
\(\mathbf{u}' = \mathbf{D}_x \cdot \mathbf{u}\). We derive
\(\mathbf{D}_x\) from Kansa 
\eqref{eq:kansa-approximation}, by linearity and
\eqref{eq:pde-kansa-linear},
\begin{equation}
\frac{\partial }{\partial x} u(x) = \sum_{k=1}^{N} \alpha_k \cdot \frac{\partial }{\partial x} \psi_k (|| x - x_k ||).
\label{eq:linear-operator-on-kansa}
\end{equation}
In matrix form, we have \(\mathbf{u}' = \mathbf{K_x} \cdot \mathbf{a}\),
where the matrix \(\mathbf{K_x} \in \mathbb{R}^{N \times N}\) is
constructed by evaluating
\(\frac{\partial }{\partial x} \psi_k (|| x - x_k ||) |_{x = x_i}\) for
all \(i, k \in \{1, \ldots, N\}\) as row and column indices.

By inverting \eqref{eq:kansa-matrix}, \(\mathbf{a} = \mathbf{K}^{-1} \cdot \mathbf{u}\), assuming \(\mathbf{K}\)
invertibility from independent basis. By
substituting \(\mathbf{a}\) into
\(\mathbf{u}' = \mathbf{K_x} \cdot \mathbf{a}\), one gets
\(\mathbf{u}' = \mathbf{K_x} \cdot \mathbf{K}^{-1} \cdot \mathbf{u}\).
The differentiable matrix is thus,
\begin{equation}
\mathbf{D}_x = \mathbf{K_x} \cdot \mathbf{K}^{-1} \in \mathbb{R}^{N \times N}.
\label{eq:differentiable-matrix}
\end{equation}
For viscous Burgers' equation \eqref{eq:burgers-general}
\(\mathcal{F}[u] = \frac{\partial u}{\partial t} + u \frac{\partial u}{\partial x} - \nu \frac{\partial^2 u}{\partial x^2}\).
Its differentiable matrix form, which follows the same formulation as in
\eqref{eq:differentiable-matrix} by replacing the operator accordingly,
is,
\begin{equation}
\mathcal{F}[u] = \mathbf{D}_t \cdot \mathbf{u} + \mathbf{u} \circ (\mathbf{D}_x \cdot \mathbf{u}) - \nu (\mathbf{D}_{xx} \cdot \mathbf{u}).
\label{eq:burgers-operator-matrix}
\end{equation}
Here we present \textbf{two} categories of Kansa approaches (Table \ref{tab:nonlinear-kansa-error-bound}). The first
consists of \textbf{four} time-stepping schemes, including two 
per-step \emph{linear} and another two \emph{nonlinear}
systems. The second employs \textbf{a fully
nonlinear solver} on the PDE residuals, without
explicit time discretization.

\begin{longtable}[]{@{}
  >{\raggedright\arraybackslash}p{(\columnwidth - 10\tabcolsep) * \real{0.1310}}
  >{\centering\arraybackslash}p{(\columnwidth - 10\tabcolsep) * \real{0.1429}}
  >{\centering\arraybackslash}p{(\columnwidth - 10\tabcolsep) * \real{0.1548}}
  >{\centering\arraybackslash}p{(\columnwidth - 10\tabcolsep) * \real{0.1429}}
  >{\centering\arraybackslash}p{(\columnwidth - 10\tabcolsep) * \real{0.2143}}
  >{\centering\arraybackslash}p{(\columnwidth - 10\tabcolsep) * \real{0.2143}}@{}}
\caption{\label{tab:nonlinear-kansa-error-bound} Summary of different
non-linear Kansa solver features, \(\Delta t\) is the time step size,
\(N_x\) and \(N_t\) are the number of collocation points in spatial and
temporal dimensions respectively.}\tabularnewline
\toprule\noalign{}
\begin{minipage}[b]{\linewidth}\raggedright
Features
\end{minipage} & \begin{minipage}[b]{\linewidth}\centering
forward
\end{minipage} & \begin{minipage}[b]{\linewidth}\centering
IMEX
\end{minipage} & \begin{minipage}[b]{\linewidth}\centering
backward
\end{minipage} & \begin{minipage}[b]{\linewidth}\centering
Crank--Nicolson
\end{minipage} & \begin{minipage}[b]{\linewidth}\centering
\textbf{fully non-linear}
\end{minipage} \\
\midrule\noalign{}
\endfirsthead
\toprule\noalign{}
\begin{minipage}[b]{\linewidth}\raggedright
Features
\end{minipage} & \begin{minipage}[b]{\linewidth}\centering
forward
\end{minipage} & \begin{minipage}[b]{\linewidth}\centering
IMEX
\end{minipage} & \begin{minipage}[b]{\linewidth}\centering
backward
\end{minipage} & \begin{minipage}[b]{\linewidth}\centering
Crank--Nicolson
\end{minipage} & \begin{minipage}[b]{\linewidth}\centering
\textbf{fully non-linear}
\end{minipage} \\
\midrule\noalign{}
\endhead
\bottomrule\noalign{}
\endlastfoot
Time-step & explicit & semi-explicit & implicit & implicit &
\(\times\) \\
Error & \(O(\Delta t)\) & \(O(\Delta t)\) & \(O(\Delta t)\) &
\(O(\Delta t^2)\) & \(O(1)\) \\
Stability & unstable & stable & stable & stable & N/A \\
Memory & \(O(N_x^2)\) & \(O(N_x^2)\) & \(O(N_x^2)\) & \(O(N_x^2)\) &
\(O(N_x^2 N_t^2)\) \\
\end{longtable}

\paragraph{Time-stepping approach with linear system.}

We can remove the non-linearity by discretizing the time derivative via
finite difference method, for a special case of time-dependent PDEs. One
solution is to use the (1) \emph{explicit forward} Euler scheme,
\begin{equation}
\frac{\partial u}{\partial t} + \mathcal{D}[u] = 0
\implies 
\frac{u^{n+1} - u^{n}}{\Delta t} + O(\Delta t) + \mathcal{D}[u^{n}] = 0,
\label{eq:nonlinear-pde-time-stepping}
\end{equation}
where \(\mathcal{D}\) is the spatial operator. A more \emph{stable}
solution is to use the (2) \emph{implicit-explicit (IMEX)} scheme, which
splits the stiff and non-stiff parts of the operator
\(\mathcal{D} = \mathcal{I}_{\text{stiff}} + \mathcal{E}_{\text{non-stiff}}\), which stiffness means the numerical instability incurred by the operator
and needs to be treated implicitly,
\begin{equation}
\frac{u^{n+1} - u^{n}}{\Delta t} + O(\Delta t)  + \mathcal{I}_{\text{stiff}}[u^{n+1}] + \mathcal{E}_{\text{non-stiff}}[u^{n}] = 0.
\label{eq:nonlinear-pde-imex}
\end{equation}
Despite the non-linear spatial operator \(\mathcal{D}\) or
\(\mathcal{E}_{\text{non-stiff}}\), we already know the solution
\(u^{n}\) at time step \(n\). Thus, with differentiable matrices, one
can evaluate \(\mathcal{D}[u^{n}]\) or
\(\mathcal{E}_{\text{non-stiff}}[u^{n}]\) directly, so as to derive the
solution \(u^{n+1}\) at the next time step \(n+1\).

\textbf{Time-stepping approach with nonlinear solver.} If we discretize the time derivative via the (3)
\emph{backward} Euler scheme, the non-linearity remains in the
formulation,
\begin{equation}
\frac{\partial u}{\partial t} + \mathcal{D}[u] = 0
\implies 
\frac{u^{n+1} - u^{n}}{\Delta t} + O(\Delta t) + \mathcal{D}[u^{n+1}]  = 0.
\label{eq:nonlinear-pde-backward-euler}
\end{equation}
We can directly replace linear system solver by a non-linear system
solver, e.g.~Newton-Raphson method \cite{Ypma1995NewtonRaphsonHistory},
to minimize the residual vector and derive the unknown solution at next
time step,
\begin{equation}
u^{n+1} = \arg \min_{u^{n+1}} \mathbf{r}^{n+1}, \text{ where } \mathbf{r}^{n+1} = u^{n+1} - u^{n} + \Delta t \cdot \mathcal{D}[u^{n+1}].
\label{eq:nonlinear-pde-residual}
\end{equation}
Alternatively, (4) \emph{Crank-Nicolson} scheme can be used to
discretize second-order accurate in time,
\begin{equation}
\frac{\partial u}{\partial t} + \mathcal{D}[u] = 0 
\implies
\frac{u^{n+1} - u^{n}}{\Delta t} + \frac{1}{2} \left( \mathcal{D}[u^{n+1}] + \mathcal{D}[u^{n}] \right) + O(\Delta t^2) = 0.
\label{eq:nonlinear-pde-crank-nicolson}
\end{equation}
Similar with \eqref{eq:nonlinear-pde-residual}, the unknown solution at
next time step is derived by minimizing the residual vector as stated in
\eqref{eq:nonlinear-pde-crank-nicolson}.


\textbf{Fully nonlinear solver without time-stepping.} This approach
directly minimizes the PDE residuals \eqref{eq:pde-kansa} over all collocation points, without explicit time discretization.
After plugging in the differentiable matrix form of the non-linear
operator \(\mathcal{F}\), the objective function is therefore,
\begin{equation}
\alpha = \arg \min_{\alpha} \sum_{i=1}^{N} \left( \mathcal{F}[\hat{u}](\mathbf{x}_i) - h(\mathbf{x}_i) \right)^2.
\label{eq:nonlinear-pde-fully-nonlinear}
\end{equation}
By plugging in Kansa approximation \eqref{eq:kansa-approximation},
we derive unknown solution \(u\) over entire domain. 


\hypertarget{auto-tuning-of-kansa-hyperparameters}{%
\subsubsection{Auto-tuning of Kansa
hyperparameters}\label{auto-tuning-of-kansa-hyperparameters}}

To tune the key Kansa method hyperparameter, kernel shape parameter \(\epsilon\) in \eqref{eq:rbf-types}, \cite{zhong2023CNF} proposed one of the self-tuning methods for \(\epsilon\) by minimizing the variation of the solution field \(u\) over all collocation points, and the condition number of operator-evaluated kernel
matrix \(\mathbf{F}\),
\begin{equation}
\epsilon^{\star} = \arg \min_{\epsilon} \omega_1 \cdot \text{cond}(\mathbf{F}) + \omega_2 \cdot \int_{\mathbb{D}} || \nabla u(\mathbf{x}) ||^2 d\mathbf{x},
\label{eq:kansa-epsilon-tuning-cond}
\end{equation}
where \(\text{cond}(\mathbf{F})\) is the condition number of matrix
\(\mathbf{F}\) defined in \eqref{eq:kansa-pde-matrix}. The integral term
can be approximated by summing over all collocation points by Monte
Carlo integration. This approach works for linear, including
\emph{coupled} and multi-dimensional, PDEs.

For \textbf{non-linear} operator case, the solution \(u\) depends on
\(\epsilon\) implicitly via the coefficients \(\alpha_i\). The matrix
\(\mathbf{F}\) no longer exists explicitly. Here, we propose to directly
minimize the PDE residuals over all collocation points, the total
variation of the solution field \(u\), and the training L2 loss between the predicted solution \(u\) and the ground truth solution \(u^{gt}\) if training data are available,
\begin{equation}
\epsilon^{\star} = \arg \min_{\epsilon} \omega_1 \cdot \sum_{i=1}^{N} \left( \mathcal{F}[\hat{u}](\mathbf{x}_i) - h(\mathbf{x}_i) \right)^2 + \omega_2 \cdot \int_{\mathbb{D}} || \nabla u(\mathbf{x}) ||^2 d\mathbf{x} + \omega_3 \cdot || u - u^{gt} ||^2,
\label{eq:kansa-epsilon-tuning-nonlinear}
\end{equation}
where \(\omega_1, \omega_2\), and \(\omega_3\) are the penalty weights. Grid search is used as an optimizer.

\hypertarget{solutions-of-inverse-pde-problems}{%
\subsection{Solutions of inverse PDE
problems}\label{solutions-of-inverse-pde-problems}}

\paragraph{Inverse PDE problems.}

When given observations of solution field \(u^{\text{obs}}\), we
infer the unknown PDE parameters \(\boldsymbol{\pi}\) that
minimize the discrepancy \(\mathcal{L}\) between the predicted
\(u^{\text{pred}}(\boldsymbol{\pi})\) and \(u^{\text{obs}}\),
\begin{equation}
\boldsymbol{\pi}^{\star} = \arg \min_{\boldsymbol{\pi}} \mathcal{L}(u^{\text{obs}}, u^{\text{pred}}(\boldsymbol{\pi})).
\label{eq:inverse-pde-objective}
\end{equation}
We adopt the \texttt{SciPy} implementation of the least squares and root finding algorithms, which are either gradient-based or gradient-free, detailed in the evaluation section.
\section{Evaluation}


\hypertarget{performance-metrics}{%
\subsection{Performance metrics}\label{performance-metrics}}

\textbf{Accuracy}. Given the numerical solution \(\hat{u}_i\) from PDE
solvers, and the ground truth \(u_i\), the \(L_2\) risk
\(\mathcal{R}_{L_2}\) is the average discretized error over all
\(N_{\text{test}}\) test points on the spatial-temporal domain,
\begin{equation}
\mathcal{\hat{R}}_{L_2} = 
\frac{1}{N_{\text{test}}}
\sum_{i=1}^{N_{\text{test}}} 
|| \hat{u}_i - u_i ||_2, 
\quad
\mathcal{\hat{R}}_{\text{relative } L_2} =
\frac{1}{N_{\text{test}}}
\sum_{i=1}^{N_{\text{test}}}
\frac{|| \hat{u}_i - u_i ||_2}{|| u_i ||_2}.
\label{eq:relative-l2-error}
\end{equation}
The relative \(L_2\) risk is computed from \(\mathcal{R}_{L_2}\) and
normalized by the ground truth \(u_i\) \(L_2\) norm,

\hypertarget{evaluation-of-solvers-for-the-advection-equation}{%
\subsection{Evaluation of solvers for the Advection
equation}\label{evaluation-of-solvers-for-the-advection-equation}}

For the 1D \textbf{advection} equation defined in
\eqref{eq:advection-general}, we set the number of domain quadrature
points \(N_{\mathcal{R}} = 100 \times 10\), i.e.~initial condition (IC)
points \(N_{d} = 10\) and the boundary condition (BC) points
\(N_{\mathcal{B}} = 100 \times 2\). The advection equation is
initialized as per Table \ref{tab:advection-1d-setup}.
\begin{longtable}[]{@{}
  >{\centering\arraybackslash}p{(\columnwidth - 8\tabcolsep) * \real{0.1875}}
  >{\centering\arraybackslash}p{(\columnwidth - 8\tabcolsep) * \real{0.1875}}
  >{\centering\arraybackslash}p{(\columnwidth - 8\tabcolsep) * \real{0.2031}}
  >{\centering\arraybackslash}p{(\columnwidth - 8\tabcolsep) * \real{0.2188}}
  >{\centering\arraybackslash}p{(\columnwidth - 8\tabcolsep) * \real{0.2031}}@{}}
\caption{\label{tab:advection-1d-setup} 1D advection equation
experimental setup.}\tabularnewline
\toprule\noalign{}
\begin{minipage}[b]{\linewidth}\centering
\textbf{domain}
\end{minipage} & \begin{minipage}[b]{\linewidth}\centering
\textbf{time range}
\end{minipage} & \begin{minipage}[b]{\linewidth}\centering
\textbf{parameter}
\end{minipage} & \begin{minipage}[b]{\linewidth}\centering
\textbf{IC}
\end{minipage} & \begin{minipage}[b]{\linewidth}\centering
\textbf{BC}
\end{minipage} \\
\midrule\noalign{}
\endfirsthead
\toprule\noalign{}
\begin{minipage}[b]{\linewidth}\centering
\textbf{domain}
\end{minipage} & \begin{minipage}[b]{\linewidth}\centering
\textbf{time range}
\end{minipage} & \begin{minipage}[b]{\linewidth}\centering
\textbf{parameter}
\end{minipage} & \begin{minipage}[b]{\linewidth}\centering
\textbf{IC}
\end{minipage} & \begin{minipage}[b]{\linewidth}\centering
\textbf{BC}
\end{minipage} \\
\midrule\noalign{}
\endhead
\bottomrule\noalign{}
\endlastfoot
\(x_0=0, x_f=1\) & \(t_0=0, t_f=1\) & \(\beta=0.4\) &
\(u_0(x) = \sin(2 \pi x)\) & per \eqref{eq:advection-solution} \\
\end{longtable}
FNO requires multiple instances of PDEs for training. Hence, we generate
\(N_{pde} = 100\) instances by varying only the initial condition as,
given \(c_k \sim \mathcal{N}(0,1)\),
\begin{equation}
u_0(x) := \frac{u_0(x)}{\max_x |u_0(x)|}, \text{ where } u_0(x) = \sum_{k=1}^{5} c_k \, \sin(2\pi k x).
\label{eq:advection-initial-condition-sampling}
\end{equation}
For training, PINN and FNO are trained via learning rate
\(\eta = 10^{-3}\) until convergence, i.e.~with epoch iterations
\(N_{\text{iter}} = 3000\) for PINN and \(N_{\text{iter}} = 100\) for
FNO. For evaluation, the test points \(N_{\text{test}} = 64 \times 8\).
The error is measured by relative \(L_2\) risk
\(\mathcal{\hat{R}}_{\text{relative } L_2}\)
\eqref{eq:relative-l2-error}.

\hypertarget{forward-problem}{%
\subsubsection{Forward problem}\label{forward-problem}}

Since FNO is trained on \(N_{pde} = 100\) instances of PDEs, we compensate more training data for single-instance solvers for a fair
comparison. The adjustment factor is defined as
\(C_{\text{scale}} \in [1, N_{pde}] \subset \mathbb{R}^+\). Hence, the
domain points is
\(N'_{\mathcal{R}} = C_{\text{scale}} \times N_{\mathcal{R}}\) and
methods denoted as \(\text{FDM}^{C_{\text{scale}}}\) and
\(\text{PINN}^{C_{\text{scale}}}\). The test-time results are summarized in Table
\ref{tab:advection-1d-error}.

\begin{longtable}[]{@{}
  >{\centering\arraybackslash}p{(\columnwidth - 8\tabcolsep) * \real{0.2041}}
  >{\centering\arraybackslash}p{(\columnwidth - 8\tabcolsep) * \real{0.1939}}
  >{\centering\arraybackslash}p{(\columnwidth - 8\tabcolsep) * \real{0.1939}}
  >{\centering\arraybackslash}p{(\columnwidth - 8\tabcolsep) * \real{0.1939}}
  >{\centering\arraybackslash}p{(\columnwidth - 8\tabcolsep) * \real{0.2143}}@{}}
\caption{\label{tab:advection-1d-error} Models accuracy
\(\mathcal{\hat{R}}_{\text{relative } L_2} \times 10^{-3}\), on 1D
advection relative to the data domain
resolution.}\tabularnewline
\toprule\noalign{}
\begin{minipage}[b]{\linewidth}\centering
\(C_{\text{scale}}\)
\end{minipage} & \begin{minipage}[b]{\linewidth}\centering
\textbf{FDM}
\end{minipage} & \begin{minipage}[b]{\linewidth}\centering
\textbf{PINN}
\end{minipage} & \begin{minipage}[b]{\linewidth}\centering
\textbf{FNO}
\end{minipage} & \begin{minipage}[b]{\linewidth}\centering
\textbf{KM}
\end{minipage} \\
\midrule\noalign{}
\endfirsthead
\toprule\noalign{}
\begin{minipage}[b]{\linewidth}\centering
\(C_{\text{scale}}\)
\end{minipage} & \begin{minipage}[b]{\linewidth}\centering
\textbf{FDM}
\end{minipage} & \begin{minipage}[b]{\linewidth}\centering
\textbf{PINN}
\end{minipage} & \begin{minipage}[b]{\linewidth}\centering
\textbf{FNO}
\end{minipage} & \begin{minipage}[b]{\linewidth}\centering
\textbf{KM}
\end{minipage} \\
\midrule\noalign{}
\endhead
\bottomrule\noalign{}
\endlastfoot
\(1\) & \(36.63\) & \(300.2\) & \(744.3\) & \(1.918\) \\
\(2^2\) & \(17.05\) & \(20.68\) & \(58.71\) & \(0.0028\) \\
\(4^2\) & \(7.478\) & \(8.654\) & \(37.68\) & N/A \\
\(N_{pde}=10^2\) & \(3.228\) & \(6.457\) & \(13.37\) & N/A \\
average & \(16.10 \pm 12.87\) & \(83.99 \pm 124.9\) &
\(213.5 \pm 306.9\) & \(0.9603 \pm 0.9576\) \\
\end{longtable}

From Table \ref{tab:advection-1d-error}, we conclude that all solvers are
sensitive to the number of training data, where larger
\(C_{\text{scale}}\) leads to better precision on test points. Kansa
outperforms other methods in both accuracy and convergence speed,
achieving the least error (up to \(10^{-6}\)) with only
\(C_{\text{scale}} = 4^2\). However, due to the increasing computational
cost above \(C_{\text{scale}} = 10^2\), memory limit was exceeded.

\hypertarget{inverse-problem}{%
\subsubsection{Inverse problem}\label{inverse-problem}}

For the 1D \textbf{advection} equation \eqref{eq:advection-general}
initialized in Table \ref{tab:advection-1d-setup}, we set up the inverse
PDE problem to infer the initial parameter \(\beta\) from the
observation data \(u^{\text{obs}}\) at all time steps. All methods are
evaluated at their best performance from the forward problem. The
results are summarized in Table \ref{tab:advection-1d-inverse-error},
with the initial parameter \(\beta_0\) set.

\begin{longtable}[]{@{}lllll lllll@{}}
\caption{\label{tab:advection-1d-inverse-error} Inverse predictions of
\(\beta\) on advection equation, where the ground truth
\(\beta = 0.4\).}\tabularnewline
\toprule\noalign{}
\(\beta_0\) & \textbf{FDM} & \textbf{PINN} & \textbf{FNO} &  \textbf{KM} & \(\beta_0\) & \textbf{FDM} & \textbf{PINN} & \textbf{FNO} &  \textbf{KM} \\
\midrule\noalign{}
\endfirsthead
\toprule\noalign{}
\(\beta_0\) & \textbf{FDM} & \textbf{PINN} & \textbf{FNO} &  \textbf{KM} & \(\beta_0\) & \textbf{FDM} & \textbf{PINN} & \textbf{FNO} &  \textbf{KM} \\
\midrule\noalign{}
\endhead
\bottomrule\noalign{}
\endlastfoot
\(0.2\) & \(0.402\) & \(0.39987\) & \(0.3985\) & \(0.402\) 
&
\(1.0\) & \(1.000\) & \(0.40446\) & \(1.2267\) & \(0.402\) \\
\end{longtable}

For local optimization methods when searching for the optimal parameter,
they stuck at different local minima depending on the initial guess
\(\beta_0\). With different runs of initial guesses, they give more
precise predictions with more computational cost.

\hypertarget{extension-1-kansa-method-for-coupled-pdes}{%
\subsection{Extension 1: Kansa method for coupled
PDEs}\label{extension-1-kansa-method-for-coupled-pdes}}

The Lotka-Volterra equations \eqref{eq:lotka-volterra} are initialized
as per Table \ref{tab:lotka-volterra-setup}, where the number of domain
quadrature points \(N_{\mathcal{R}} = 100 \times 1\), and initial condition
points \(N_{d} = 1\). For evaluation, the test points
\(N_{\text{test}} = 64\). The results from Kansa method are summarized
in Table \ref{tab:LV-maxwell-error}, where the Gaussian RBF shape
parameters, as defined in \eqref{eq:rbf-types}, are set as
\(\epsilon = 0.2\) for both \(x(t)\) and \(y(t)\).

\begin{longtable}[]{@{}
  >{\centering\arraybackslash}p{(\columnwidth - 4\tabcolsep) * \real{0.2041}}
  >{\centering\arraybackslash}p{(\columnwidth - 4\tabcolsep) * \real{0.5204}}
  >{\centering\arraybackslash}p{(\columnwidth - 4\tabcolsep) * \real{0.2755}}@{}}
\caption{\label{tab:lotka-volterra-setup} 1D Lotka-Volterra equations
experimental setup.}\tabularnewline
\toprule\noalign{}
\begin{minipage}[b]{\linewidth}\centering
\textbf{time range}
\end{minipage} & \begin{minipage}[b]{\linewidth}\centering
\textbf{parameter}
\end{minipage} & \begin{minipage}[b]{\linewidth}\centering
\textbf{initial conditions}
\end{minipage} \\
\midrule\noalign{}
\endfirsthead
\toprule\noalign{}
\begin{minipage}[b]{\linewidth}\centering
\textbf{time range}
\end{minipage} & \begin{minipage}[b]{\linewidth}\centering
\textbf{parameter}
\end{minipage} & \begin{minipage}[b]{\linewidth}\centering
\textbf{initial conditions}
\end{minipage} \\
\midrule\noalign{}
\endhead
\bottomrule\noalign{}
\endlastfoot
\(t_0=0, t_f=200\) & \(\alpha=0.1, \beta=0.02, \delta=0.01, \gamma=0.1\)
& \(x(0) = 40, y(0) = 9\) \\
\end{longtable}

The 1D \textbf{Maxwell's} equations as defined in
\eqref{eq:maxwell-1d-wave} are initialized per Table
\ref{tab:maxwell-1d-setup}, where the speed of time propagation \(c=1\),
the number of domain quadrature points
\(N_{\mathcal{R}} = 12 \times 12\), and initial condition points
\(N_{d} = 24\). For evaluation, the test points
\(N_{\text{test}} = 10 \times 10\). The shape parameter of Gaussian RBF,
as defined in \eqref{eq:rbf-types}, is set as \(\epsilon_x = 0.21\) and
\(\epsilon_y = 0.2\) for Lotka-Volterra equations and
\(\epsilon_E = 16\) and \(\epsilon_B = 16\) for Maxwell's equations,
respectively.

\hypertarget{forward-problem-1}{%
\subsubsection{Forward problem}\label{forward-problem-1}}

\begin{longtable}[]{@{}
  >{\centering\arraybackslash}p{(\columnwidth - 8\tabcolsep) * \real{0.2073}}
  >{\centering\arraybackslash}p{(\columnwidth - 8\tabcolsep) * \real{0.1707}}
  >{\centering\arraybackslash}p{(\columnwidth - 8\tabcolsep) * \real{0.1829}}
  >{\centering\arraybackslash}p{(\columnwidth - 8\tabcolsep) * \real{0.2195}}
  >{\centering\arraybackslash}p{(\columnwidth - 8\tabcolsep) * \real{0.2195}}@{}}
\caption{\label{tab:LV-maxwell-error}
\(\mathcal{\hat{R}}_{\text{relative } L_2}\) error of Lotka-Volterra and
Maxwell's equations using Kansa method.}\tabularnewline
\toprule\noalign{}
\begin{minipage}[b]{\linewidth}\centering
\(C_{\text{scale}}\)
\end{minipage} & \begin{minipage}[b]{\linewidth}\centering
\(x(t)\)
\end{minipage} & \begin{minipage}[b]{\linewidth}\centering
\(y(t)\)
\end{minipage} & \begin{minipage}[b]{\linewidth}\centering
\(E_z(x, t)\)
\end{minipage} & \begin{minipage}[b]{\linewidth}\centering
\(B_y(x, t)\)
\end{minipage} \\
\midrule\noalign{}
\endfirsthead
\toprule\noalign{}
\begin{minipage}[b]{\linewidth}\centering
\(C_{\text{scale}}\)
\end{minipage} & \begin{minipage}[b]{\linewidth}\centering
\(x(t)\)
\end{minipage} & \begin{minipage}[b]{\linewidth}\centering
\(y(t)\)
\end{minipage} & \begin{minipage}[b]{\linewidth}\centering
\(E_z(x, t)\)
\end{minipage} & \begin{minipage}[b]{\linewidth}\centering
\(B_y(x, t)\)
\end{minipage} \\
\midrule\noalign{}
\endhead
\bottomrule\noalign{}
\endlastfoot
\(1\) & \(0.1279353\) & \(0.055667494\) & 0.8049189 & 0.5894967 \\
\(4\) & \(0.04539858\) & \(0.06230465\) & 0.4383743 & 0.3830594 \\
\end{longtable}

\textbf{Accuracy}. The results from Kansa method are summarized in Table
\ref{tab:LV-maxwell-error}. Both errors converge with
increasing \(C_{\text{scale}}\) as defined above. \textbf{Efficiency}. The training time and inference time of Kansa method on Lotka-Volterra equations are \(0.4034\) and \(0.0001\) seconds, respectively. The training time and inference time
of Kansa method on Maxwell's equations are \(0.4486\) and
\(0.0005\) seconds.

\hypertarget{inverse-problem-1}{%
\subsubsection{Inverse problem}\label{inverse-problem-1}}

For the Lotka-Volterra defined in \eqref{eq:lotka-volterra}
initialized in Table \ref{tab:lotka-volterra-setup}, we set up the
inverse problem to infer the initial parameter \(\alpha\),
\(\beta\), \(\delta\) and \(\gamma\) from observation
\(x^{\text{obs}}(t)\) and \(y^{\text{obs}}(t)\) at all time steps.


\begin{longtable}[]{@{}lcccc lcccc@{}}
\caption{\label{tab:lotka-volterra-inverse-acc} Inverse predictions of
\(\alpha\), \(\beta\), \(\delta\) and \(\gamma\) on Lotka-Volterra
equations.}\tabularnewline
\toprule\noalign{}
& \(\alpha\) & \(\beta\) & \(\delta\) & \(\gamma\) 
&
& \(\alpha\) & \(\beta\) & \(\delta\) & \(\gamma\) \\
\midrule\noalign{}
\endfirsthead
\toprule\noalign{}
& \(\alpha\) & \(\beta\) & \(\delta\) & \(\gamma\) & 
& \(\alpha\) & \(\beta\) & \(\delta\) & \(\gamma\)  \\
\midrule\noalign{}
\endhead
\bottomrule\noalign{}
\endlastfoot
reference & \(0.1\) & \(0.02\) & \(0.01\) & \(0.1\) &
prediction & \(0.102\) & \(0.0207\) & \(0.0100\) & \(0.0994\) \\
\end{longtable}

With the initial guess all set to \(1\), the results are summarized in
Table \ref{tab:lotka-volterra-inverse-acc}. Despite the four-dimensional
search space, the optimization algorithm \texttt{SciPy} Powell method
successfully infers the parameters with high accuracy and decent
computational cost.

\hypertarget{extension-2-kansa-method-for-nonlinear-pdes}{%
\subsection{Extension 2: Kansa method for nonlinear
PDEs}\label{extension-2-kansa-method-for-nonlinear-pdes}}

The \textbf{Burgers'} equation defined in \eqref{eq:burgers-general} is
initialized as per Table \ref{tab:burgers-setup}, where the number of
domain quadrature points \(N_{\mathcal{R}} = 64 \times 16\),
i.e.~initial condition (IC) points \(N_{d} = 64\) and the boundary
condition (BC) points \(N_{\mathcal{B}} = 16 \times 2\). For evaluation,
the test points \(N_{\text{test}} = 48 \times 12\). The Gaussian RBF shape parameter, as
defined in \eqref{eq:rbf-types}, is set as \(\epsilon = 0.9\).

\begin{longtable}[]{@{}
  >{\centering\arraybackslash}p{(\columnwidth - 8\tabcolsep) * \real{0.23}}
  >{\centering\arraybackslash}p{(\columnwidth - 8\tabcolsep) * \real{0.2}}
  >{\centering\arraybackslash}p{(\columnwidth - 8\tabcolsep) * \real{0.10}}
  >{\centering\arraybackslash}p{(\columnwidth - 8\tabcolsep) * \real{0.18}}
  >{\centering\arraybackslash}p{(\columnwidth - 8\tabcolsep) * \real{0.30}}@{}}
\caption{\label{tab:burgers-setup} Burgers' equation experimental
setup.}\tabularnewline
\toprule\noalign{}
\begin{minipage}[b]{\linewidth}\centering
\textbf{domain}
\end{minipage} & \begin{minipage}[b]{\linewidth}\centering
\textbf{time span}
\end{minipage} & \begin{minipage}[b]{\linewidth}\centering
\textbf{param.}
\end{minipage} & \begin{minipage}[b]{\linewidth}\centering
\textbf{ICs}
\end{minipage} & \begin{minipage}[b]{\linewidth}\centering
\textbf{BCs}
\end{minipage} \\
\midrule\noalign{}
\endfirsthead
\toprule\noalign{}
\begin{minipage}[b]{\linewidth}\centering
\textbf{domain}
\end{minipage} & \begin{minipage}[b]{\linewidth}\centering
\textbf{time span}
\end{minipage} & \begin{minipage}[b]{\linewidth}\centering
\textbf{parameter}
\end{minipage} & \begin{minipage}[b]{\linewidth}\centering
\textbf{ICs}
\end{minipage} & \begin{minipage}[b]{\linewidth}\centering
\textbf{BCs}
\end{minipage} \\
\midrule\noalign{}
\endhead
\bottomrule\noalign{}
\endlastfoot
\(x_0=-10, x_f=10\) & \(t_0=0, t_f=4\) & \(\nu=0.5\) & per
\eqref{eq:burgers-sol-steady} & \(u(x_0) = 1, u(x_f) = 0\) \\
\end{longtable}

\hypertarget{forward-problem-2}{%
\subsubsection{Forward problem}\label{forward-problem-2}}

\textbf{Accuracy}. From Table \ref{tab:burgers-error}, we observe that
fully non-linear approach outperforms other time-stepping schemes. It's hard to determine whether IMEX or backward Euler
is more accurate theoretically. However, Crank-Nicolson scheme is definitely more accurate than both IMEX and backward
Euler, since it's second-order accurate in time while the other two are
only first-order accurate.

\begin{longtable}[]{@{}ccccc@{}}
\caption{\label{tab:burgers-error}
\(\mathcal{\hat{R}}_{\text{relative } L_2} \times 10^{-2}\) error of
Burgers' equation using Kansa methods.}\tabularnewline
\toprule\noalign{}
forward & IMEX & backward & Crank--Nicolson & \textbf{fully
non-linear} \\
\midrule\noalign{}
\endfirsthead
\toprule\noalign{}
forward & IMEX & backward & Crank--Nicolson & \textbf{fully
non-linear} \\
\midrule\noalign{}
\endhead
\bottomrule\noalign{}
\endlastfoot
\(3.74 \times 10^{31}\) & \(1.68\) & \(1.33\) & \(1.29\) & \(0.012\) \\
\end{longtable}

\textbf{Computational efficiency}. We measure the training and
inference time of different Kansa methods on Burgers' equation in Table \ref{tab:burgers-time}. The non-linear solver used is the \texttt{SciPy} least-squares.
\begin{longtable}[]{@{}cccccc@{}}
\caption{\label{tab:burgers-time} Train or infer time of Burgers'
equation using Kansa methods (in seconds).}\tabularnewline
\toprule\noalign{}
& forward & IMEX & backward & Crank--Nicolson & \emph{fully non-linear} \\
\midrule\noalign{}
\endfirsthead
\toprule\noalign{}
forward & IMEX & backward & Crank--Nicolson & \emph{fully non-linear} \\
\midrule\noalign{}
\endhead
\bottomrule\noalign{}
\endlastfoot
Training &\(0.34\) & \(0.47\) & \(2.33\) & \(1.66\) & \(99.2\) \\
Inference &\(0.436\) & \(0.272\) & \(1.053\) & \(1.109\) & \(0.005\) \\
\end{longtable}
\textbf{Training} time for fully nonlinear approach is longer because each step involves heavier computation with substantial memory (Table
\ref{tab:nonlinear-kansa-error-bound}), further compounded by nonlinear solvers. Four time-stepping schemes have
much less training time. The forward Euler is unstable when the stability condition is not satisfied. \textbf{Inference} time of fully
non-linear approach is significantly reduced, due to the reuse of
coefficient from the training phase. Despite a full test-time recomputation from scratch, the inference time of four time-stepping schemes remains acceptable for most practical applications.



\hypertarget{inverse-problem-2}{%
\subsubsection{Inverse problem}\label{inverse-problem-2}}

For the Burgers' equation defined in \eqref{eq:burgers-general}
initialized in Table \ref{tab:burgers-setup}, we set up the inverse PDE
problem to infer the initial parameter \(\nu\) from the observation data
\(u^{\text{obs}}\) at all time steps. The results are summarized in
Table \ref{tab:burger-inverse-acc}, with the initial parameter \(\nu_0\)
set as \(0.1\).

\begin{longtable}[]{@{}
  >{\centering\arraybackslash}p{(\columnwidth - 8\tabcolsep) * \real{0.2000}}
  >{\centering\arraybackslash}p{(\columnwidth - 8\tabcolsep) * \real{0.2000}}
  >{\centering\arraybackslash}p{(\columnwidth - 8\tabcolsep) * \real{0.2000}}
  >{\centering\arraybackslash}p{(\columnwidth - 8\tabcolsep) * \real{0.2000}}
  >{\centering\arraybackslash}p{(\columnwidth - 8\tabcolsep) * \real{0.2000}}@{}}
\caption{\label{tab:burger-inverse-acc} Inverse predictions of \(\nu\)
on Burgers' equation, where the ground truth
\(\nu = 0.5\).}\tabularnewline
\toprule\noalign{}
\begin{minipage}[b]{\linewidth}\centering
forward
\end{minipage} & \begin{minipage}[b]{\linewidth}\centering
IMEX
\end{minipage} & \begin{minipage}[b]{\linewidth}\centering
backward
\end{minipage} & \begin{minipage}[b]{\linewidth}\centering
\textbf{Crank--Nicolson}
\end{minipage} & \begin{minipage}[b]{\linewidth}\centering
\textbf{\emph{fully non-linear}}
\end{minipage} \\
\midrule\noalign{}
\endfirsthead
\toprule\noalign{}
\begin{minipage}[b]{\linewidth}\centering
forward
\end{minipage} & \begin{minipage}[b]{\linewidth}\centering
IMEX
\end{minipage} & \begin{minipage}[b]{\linewidth}\centering
backward
\end{minipage} & \begin{minipage}[b]{\linewidth}\centering
\textbf{Crank--Nicolson}
\end{minipage} & \begin{minipage}[b]{\linewidth}\centering
\textbf{\emph{fully non-linear}}
\end{minipage} \\
\midrule\noalign{}
\endhead
\bottomrule\noalign{}
\endlastfoot
\(0.388\) & \(0.535\) & \(0.467\) & \(0.502\) & \(0.500\) \\
\end{longtable}

\textbf{Accuracy}. Under same optimizer and initial guess,
the Crank-Nicolson scheme confirms its theoretical advantage (Table
\ref{tab:nonlinear-kansa-error-bound}) over both IMEX and backward
Euler, which stuck at local minima. \textbf{Computational
efficiency}. Fully non-linear approach requires retraining for each new
parameter, which is computationally expensive (Table
\ref{tab:burgers-time}). To speed up the per-run training time, it
is trained with a maximum iteration. \textbf{Stability}. The forward Euler scheme is unstable when given large \(\Delta t\). 

\section{Conclusions}

This paper extends \cite{zhong2023CNF} RBF framework solver beyond original scope of linear PDEs. In particular, we generalize its PDE solver to handle \textbf{coupled} and \textbf{nonlinear} PDEs, addressing the loss property of linear reordering. These broaden the applicability of CNF-driven self-tuning (Appendix \ref{auto-tuning-of-kansa-hyperparameters}) mesh-free solvers to both forward modeling and inverse problem formulations.

In addition, this work contributes a systematic empirical study of how CNF solvers compare with established classical and neural PDE solvers. By implementing representative prior methods and evaluating them across benchmark problems, we assess their relative performance in terms of solution accuracy, efficiency, convergence and complexity. Such comparisons clarify the strengths and limitations of CNF-based approaches within the broader landscape of PDE solvers.

Overall, this paper demonstrates that learning-guided Kansa solvers can serve as a promising and flexible tool for coupled or nonlinear PDE systems. \textbf{Future work} includes theoretical analysis of error and convergence properties, application to neural field in computing, and integration with differentiable pipelines in scientific domains.



\newpage
\bibliography{main}
\bibliographystyle{iclr2026_conference}

\newpage

\appendix

\section{Linear operator}\label{linear-operator}

A \textbf{linear operator} \cite{Iserles2008NumericalAnalysis} is a
function \(\mathcal{F}: V \rightarrow W\) that maps one vector space
\(V \in \mathbb{R}\) to another, or
itself\footnote{If the domain and codomain are the same vector space, i.e., $\mathcal{F}: V \rightarrow V$, it’s called a linear transformation or operator on $V$.},
\(W \in \mathbb{R}\), and preserving the operations of \textbf{vector
addition} and \textbf{scalar multiplication}, also known as
\textbf{homogeneity}. Thus, for all vectors \(\mathbf{u_i} \in V\) and
all scalars \(c\), the following features hold:
\begin{equation}
\begin{aligned}
\mathcal{F}(\sum_{i=1}^n \mathbf{u}_i) = \sum_{i=1}^n \mathcal{F}(\mathbf{u}_i), \quad \text{vector additivity}, \\
\mathcal{F}(c \cdot \mathbf{u}) = c \cdot\mathcal{F}(\mathbf{u}), \quad \text{scalar multiplication}.
\end{aligned}
\end{equation}
Linear operators are fundamental in Linear Algebra for processing
matrices, Quantum Mechanics for observables, Machine Learning, and
Signal Processing. This forms the basis for Kansa method
for linear PDEs.

Here are several commonly used examples of linear operators below, among which some are used in this work for PDE solver algorithms.

\begin{itemize}
\tightlist
\item
  \textbf{Matrix multiplication}: For a matrix \(A\), the function
  \(A: \mathbb{R}^n \rightarrow \mathbb{R}^m\) is a linear operator,
\end{itemize}
\begin{equation}
\mathcal{F}(\mathbf{x}) = A\mathbf{x}.
\end{equation}
\begin{itemize}
\tightlist
\item
  \textbf{Integral operator}: The operator that integrates a function
  over a fixed interval \([a, b]\) is a linear operator,
\end{itemize}
\begin{equation}
I(f) = \int_a^b f(x)\, dx.
\end{equation}
\begin{itemize}
\tightlist
\item
  \textbf{Differentiation}: The operator taking the derivative in a
  function space is a linear operator, because differentiation preserves
  addition and scalar multiplication,
\end{itemize}
\begin{equation}
D_x(f) = \frac{\partial f}{\partial x}.
\end{equation}
\begin{itemize}
\tightlist
\item
  \textbf{Gradient operator}: In multivariable calculus, the gradient
  operator \(\nabla\) is a linear operator that maps a scalar field to a
  vector field,
\end{itemize}
\begin{equation}
\nabla f = \left( \frac{\partial f}{\partial x}, \frac{\partial f}{\partial y}, \frac{\partial f}{\partial z} \right).
\end{equation}
\begin{itemize}
\tightlist
\item
  \textbf{Divergence operator}: In vector calculus, the divergence
  operator \(\nabla \cdot\) is a linear operator that maps a vector
  field to a scalar field,
\end{itemize}
\begin{equation}
\nabla \cdot \mathbf{F} = \frac{\partial F_x}{\partial x} + \frac{\partial F_y}{\partial y} + \frac{\partial F_z}{\partial z}.
\end{equation}
\begin{itemize}
\tightlist
\item
  \textbf{Laplace operator}: In the context of partial differential
  equations, the Laplace operator \(\Delta\) is a linear operator that
  maps a scalar field to another scalar field,
\end{itemize}
\begin{equation}
\Delta f = \nabla  \cdot (\nabla f) = \nabla^2 f = \frac{\partial^2 f}{\partial x^2} + \frac{\partial^2 f}{\partial y^2} + \frac{\partial^2 f}{\partial z^2}.
\end{equation}
\begin{itemize}
\tightlist
\item
  \textbf{Curl operator}: In vector calculus, the curl operator
  \(\nabla \times\) is a linear operator that maps a vector field to
  another vector field,
\end{itemize}
\begin{equation}
\begin{aligned}
\nabla \times \mathbf{F} 
&= \left( \frac{\partial F_z}{\partial y} - \frac{\partial F_y}{\partial z}, \frac{\partial F_x}{\partial z} - \frac{\partial F_z}{\partial x}, \frac{\partial F_y}{\partial x} - \frac{\partial F_x}{\partial y} \right)
&= 
\begin{vmatrix}
\hat{i} & \hat{j} & \hat{k} \\
\frac{\partial}{\partial x} & \frac{\partial}{\partial y} & \frac{\partial}{\partial z} \\
F_x & F_y & F_z
\end{vmatrix}.
\end{aligned}
\end{equation}

\hypertarget{partial-differential-equations}{%
\section{Partial differential
equations}\label{partial-differential-equations}}

\hypertarget{boundary-and-initial-conditions-bcs-and-ics}{%
\subsection{Boundary and initial conditions (BCs and
ICs)}\label{boundary-and-initial-conditions-bcs-and-ics}}

Since solution to differential equations contain integration constants,
which is non-unique, additional conditions are required to enforce
uniqueness. The boundary conditions (BCs) specify the function \(u\)
behavior on the domain boundary \(\partial \Omega\), whereas the initial
conditions (ICs) from time scale perspective are given at \(t=0\). The
formulation is defined in \eqref{eq:pde-general}.

There are some common boundary conditions, defined over the boundary
\(\Omega = [x_0, x_f]\) in 1D space, where \(\{g_i\}_{i=1}^4\) are given
closed-form functions,
\begin{equation}
\begin{aligned}
\text{Zero BC:} \quad & u(x_0, t) = 0, \; u(x_f, t) = 0, \\
\text{Dirichlet BC:} \quad & u(x_0, t) = g_1(t), \; u(x_f, t) = g_2(t), \\
\text{von Neumann BC:} \quad & \frac{\partial u}{\partial x}(x_0, t) = g_3(t), \; \frac{\partial u}{\partial x}(x_f, t) = g_4(t). \\
\end{aligned}
\label{eq:pde-bc}
\end{equation}

\subsection{Summary of PDEs}\label{summary-of-pdes}

\begin{table}[h!]
\centering
\caption{Summary of PDEs with different characteristics.}
\begin{tabular}{l l l l}
\toprule
\textbf{Equation} & \textbf{Domains} & \textbf{Linearity} & \textbf{Solution dim.} \\
\midrule
Advection & Physics, Graphics & Linear & 1 \\
Wave & Physics, Graphics & Linear & 1 \\
Lotka-Volterra & Biology & Linear & \textbf{2} \\
Maxwell & Physics & Linear & \textbf{2} \\
Burgers & Physics, Graphics & \textbf{Nonlinear} & 1 \\
\bottomrule
\end{tabular}
\label{tab:pde_summary}
\end{table}

\hypertarget{d-advection-equation}{%
\subsection{1D Advection equation}\label{d-advection-equation}}

\begin{figure}[htpb]
    \centering
    \begin{subfigure}{0.36\textwidth}
        \centering
        \includegraphics[width=\textwidth]{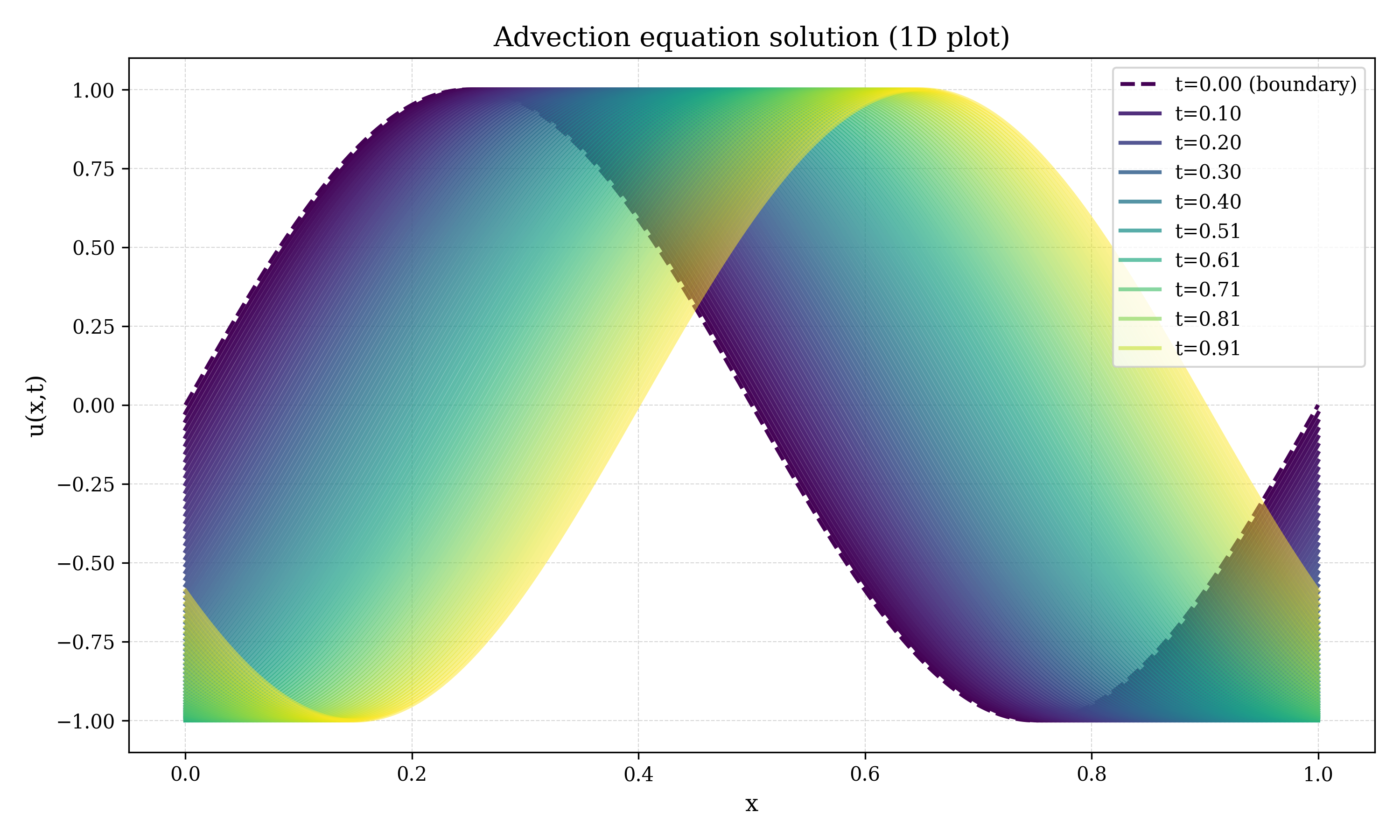}
        \label{fig:advection-solution}
    \end{subfigure}
    \hfill
    \begin{subfigure}{0.36\textwidth}
        \centering
        \includegraphics[width=\textwidth]{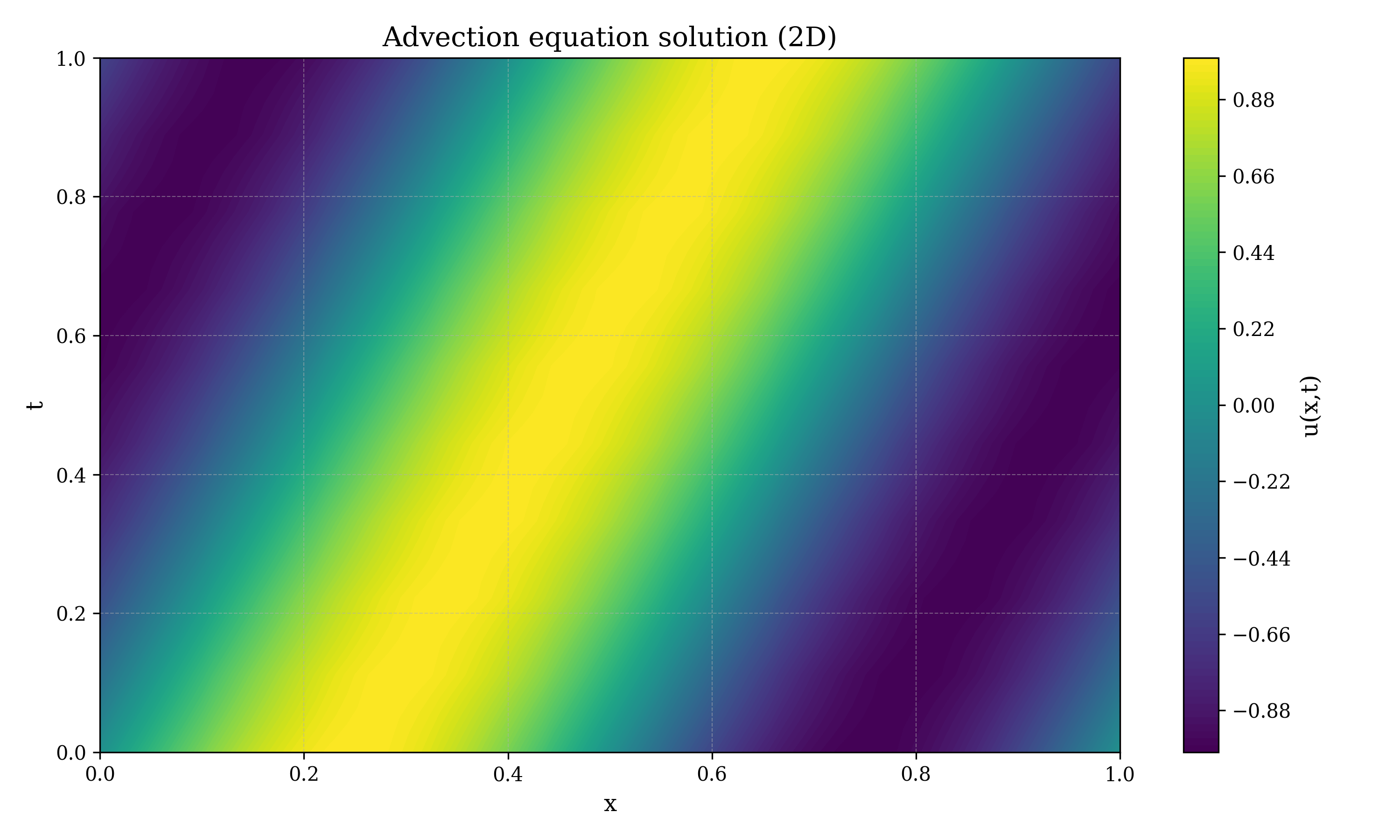}
        \label{fig:advection-solution-colormap}
    \end{subfigure}
    \hfill
        \begin{subfigure}{0.26\textwidth}
        \centering
        \includegraphics[width=\textwidth]{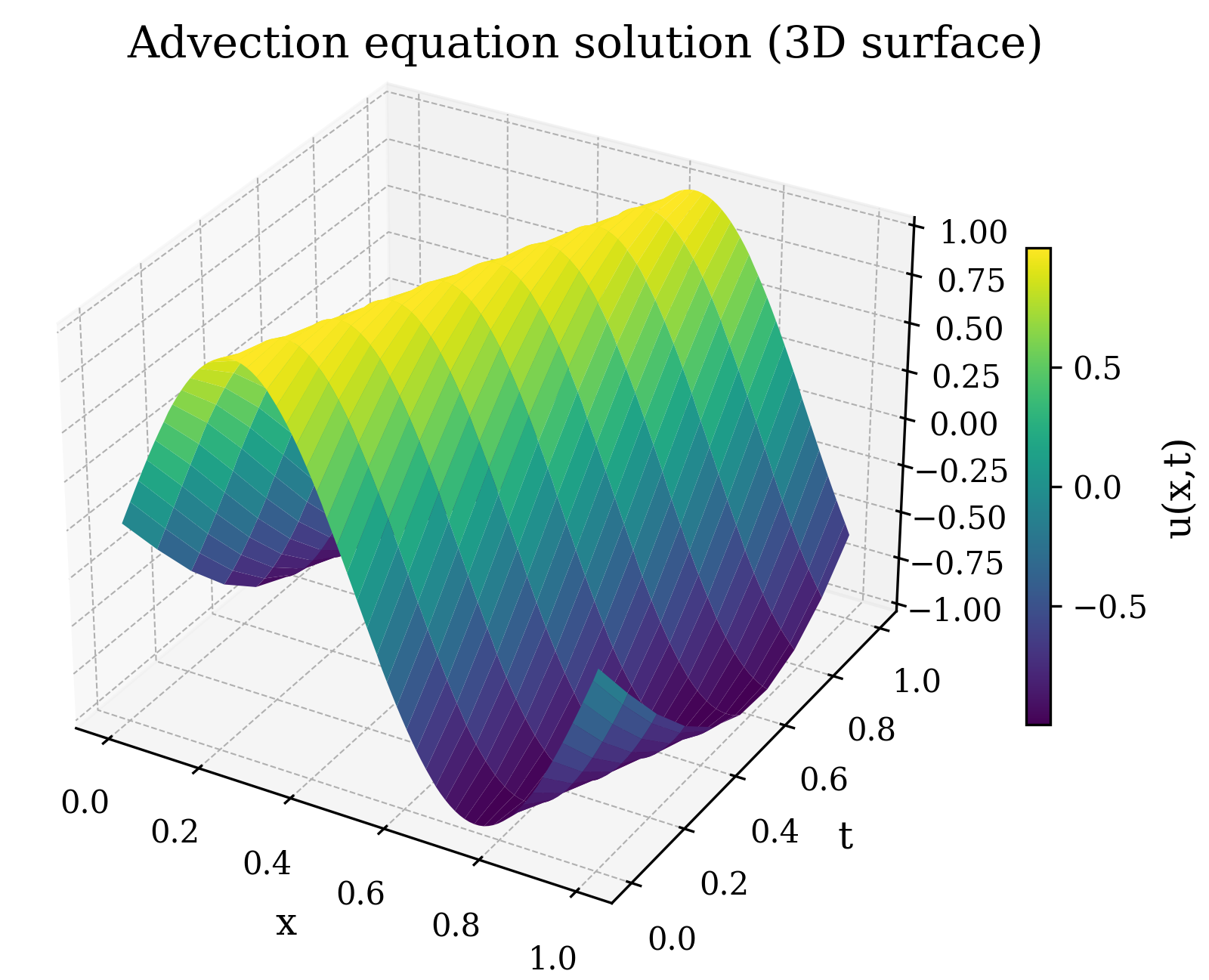}
        \label{fig:advection-solution-3d}
    \end{subfigure}
    \caption{Advection equation solution visualization in 1D, 2D and 3D.}
    \label{fig:advection-solution-visualization}
\end{figure}

The advection equation \cite{Takamoto2022PDEBench} 
models the linear transport of a scalar quantity \(u(x, t)\), which is
changed over time \(t\) and space \(x\), as follows:
\begin{equation}
\begin{aligned}
\begin{cases}
\frac{\partial u(x, t)}{\partial t} + \beta \frac{\partial u(x, t)}{\partial x} = 0, \quad x \in [x_0, x_f], t \in [t_0, t_f], \\
u(x, 0) = u_0(x), \quad x \in [x_0, x_f], \quad \text{initial condition},
\end{cases}
\end{aligned}
\label{eq:advection-general}
\end{equation}
where parameter \(\beta \in \mathbb{R}\) is the advection velocity, and
\(u_0(x)\) is the initial condition given at \(t=0\). The analytical
solution of \eqref{eq:advection-general} is,
\begin{equation}
u(x, t) = u_0(x - \beta t).
\label{eq:advection-solution}
\end{equation}
The positivity of parameter \(\beta\) indicates the direction of wave
propagation. From \eqref{eq:advection-solution}, when \(\beta > 0\), the
wave propagates rightwards, and vice versa. The solution is visualized
in Figure \ref{fig:advection-solution-visualization}, with initial
condition \(u_0(x) = \sin(2 \pi x)\), \(x \in [0, 1]\).

\hypertarget{lotka-volterra-predator-prey-model}{%
\subsection{Lotka-Volterra predator-prey
model}\label{lotka-volterra-predator-prey-model}}

Lotka-Volterra predator-prey model \cite{bacaer2011lotka} relates the
populations of prey \(x(t)\) and predators \(y(t)\) at time \(t\) in a
dynamic biological system via coupled differential equations, also
applicable to other fields, e.g.~the unemployment rate with respect to
wage growth \cite{orlando2021growth} and many more,
\begin{equation}
\begin{cases}
x'(t):= \frac{d x(t)}{dt} = \alpha x(t) - \beta x(t) \cdot y(t), \\
y'(t):= \frac{d y(t)}{dt} = \delta x(t) \cdot y(t) - \gamma y(t).
\end{cases}, t \in [t_0, t_f].
\label{eq:lotka-volterra}
\end{equation}
where \(\alpha\) is the prey growth rate, \(\beta\) is the predation
rate, \(\delta\) is the ratio of neonate predators to eaten prey, and
\(\gamma\) is the predator death rate. It assumes that there would be
unlimited food supply for the prey, and thus exponential growth
\(\alpha x (t)\). The multiplicative term \(x(t) \cdot y(t)\) represents
the encounters between prey and predators statistically.

\begin{figure}[htpb]
    \centering
    \begin{subfigure}{0.45\textwidth}
        \centering
        \includegraphics[width=\textwidth]{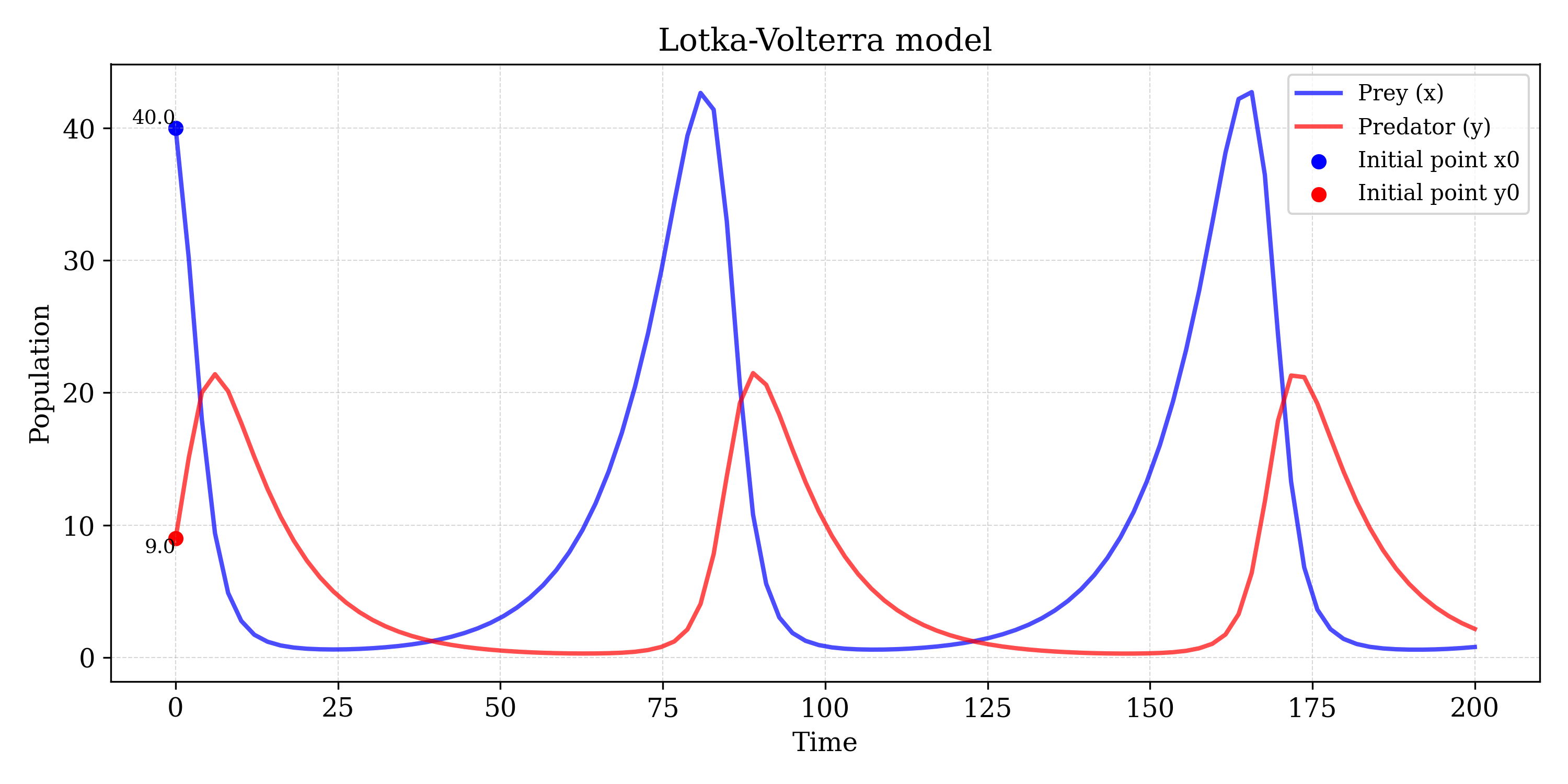}
        \caption{Prey and predator solution populations}
        \label{fig:lotka-volterra-solution}
    \end{subfigure}
    \hfill
    \begin{subfigure}{0.45\textwidth}
        \centering
        \includegraphics[width=\textwidth]{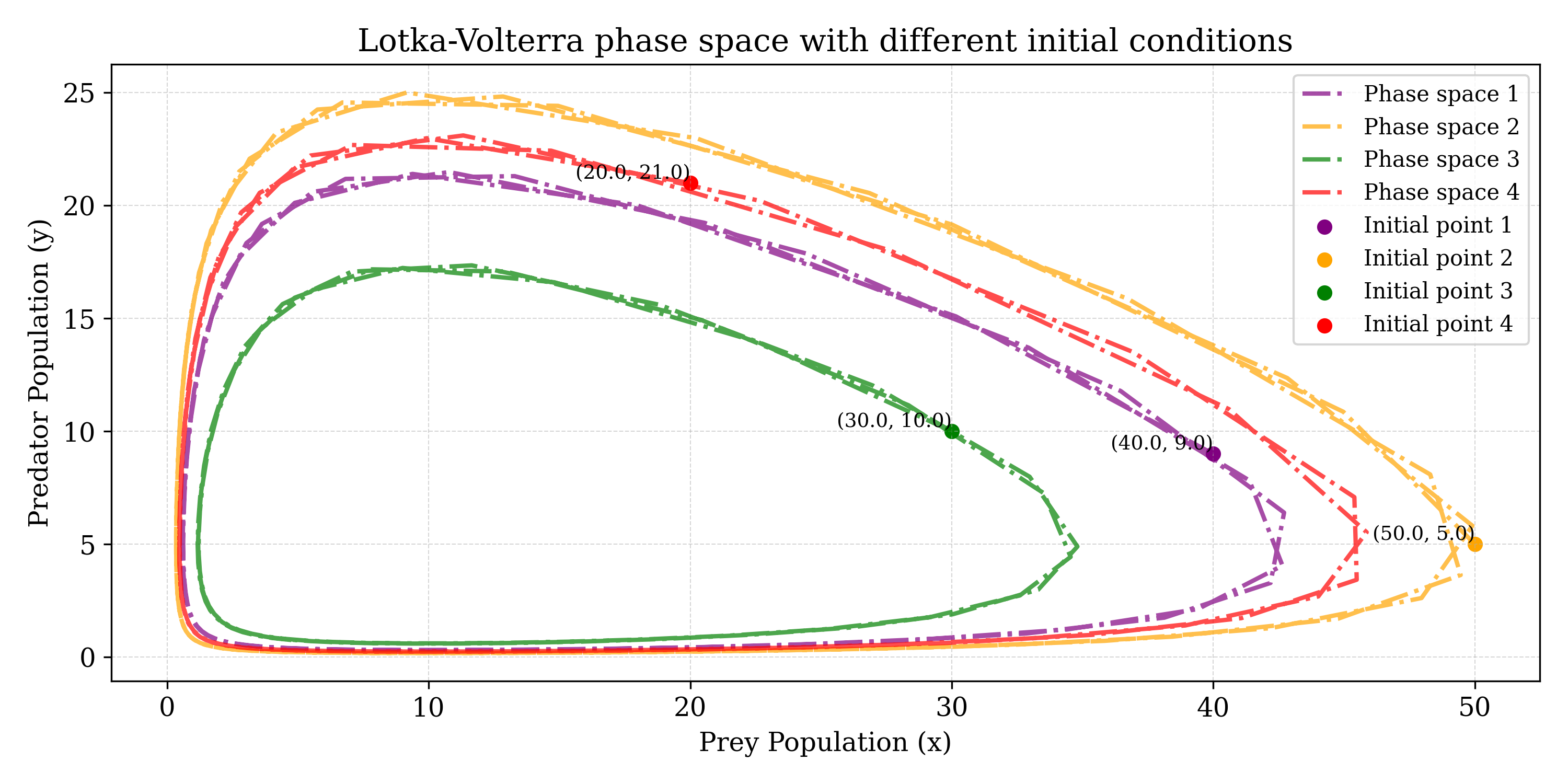}
        \caption{Phase space trajectory}
        \label{fig:lotka-volterra-phase}
    \end{subfigure}
    \caption{Lotka-Volterra predator-prey model solution and phase space.}
    \label{fig:lotka-volterra-visualization}
\end{figure}

The system has no explicit analytical solution, but the implicit
solution exists. After scaling of variables,
\begin{equation}
x^{\ast}(t) = \frac{\delta}{\gamma} x(t), \quad y^{\ast}(t) = \frac{\beta}{\alpha} y(t), \quad \tau = \alpha t,
\end{equation}
By plugging into \eqref{eq:lotka-volterra}, and dividing the first
equation by the second,
\begin{equation}
\frac{d y^{\ast}}{d x^{\ast}} = \frac{\gamma}{\alpha} \cdot \frac{y^{\ast} (x^{\ast} - 1)}{x^{\ast} (y^{\ast} - 1)},
\end{equation}
The implicit solution is given by integration separation of variables,
for which \(C_{\text{L-V}} \in \mathbb{R}\) is the integration constant,
\begin{equation}
\ln(y^{\ast}) - y^{\ast} - \frac{\gamma}{\alpha} [\ln(x^{\ast}) - x^{\ast}] = C_{\text{L-V}}.
\end{equation}
Figure \ref{fig:lotka-volterra-visualization} shows the solution with
phase space given by the above implicit solution, which depends on the
initial conditions \(x(0)\) and \(y(0)\).

\hypertarget{maxwells-equations}{%
\subsection{Maxwell's equations}\label{maxwells-equations}}

In electromagnetism, Maxwell's equations
\cite{Greiner1998MaxwellsEquations} relate the electric field
\(\mathbf{E}(\mathbf{r}, t)\) and magnetic field
\(\mathbf{B}(\mathbf{r}, t)\) with spatial position
\(\mathbf{r} \in \mathbb{R}^3\) and time \(t\), to the electric charge
density \(\rho(\mathbf{r}, t) \in \mathbb{R}\) and current density
\(\mathbf{J}(\mathbf{r}, t) \in \mathbb{R}^3\). The differential form is
as follows,
\begin{equation}
\begin{aligned}
\begin{cases}
\nabla \cdot \mathbf{E} = \frac{\rho}{\epsilon_0},& \quad \text{Gauss's law}, \\
\nabla \cdot \mathbf{B} = 0, & \quad \text{Gauss's law for magnetism}, \\
\nabla \times \mathbf{E} = - \frac{\partial \mathbf{B}}{\partial t}, & \quad \text{Faraday's law of induction}, \\  
\nabla \times \mathbf{B} = \mu_0 \mathbf{J} + \mu_0 \epsilon_0 \frac{\partial \mathbf{E}}{\partial t}, & \quad \text{Ampère-Maxwell law},
\end{cases}
\end{aligned}
\label{eq:maxwell-differential}
\end{equation}
where constants \(\mu_0, \epsilon_0 \in \mathbb{R}^+\) are the vacuum
permeability and permittivity respectively. Their product is the
reciprocal of the square of the speed of light
\(c \approx 3 \times 10^8\) \(\mathrm{m\,s^{-1}}\) in vacuum,
\begin{equation}
\mu_0 \epsilon_0 = \frac{1}{c^2}.
\label{eq:mu-epsilon-c}
\end{equation}
The first two equations state that the electric field \(\mathbf{E}\)
sourced by electric charges, and no magnetic monopoles exist. The last
two equations depict how a time-varying magnetic field \(\mathbf{B}\)
induces an electric field \(\mathbf{E}\), and vice versa with the
addition of current density \(\mathbf{J}\). In general, Maxwell's
equations are \emph{linear} with respect to \(\mathbf{E}\) and
\(\mathbf{B}\).

\begin{figure}[htpb]
    \centering
    \begin{subfigure}{0.36\textwidth}
        \centering
        \includegraphics[width=\textwidth]{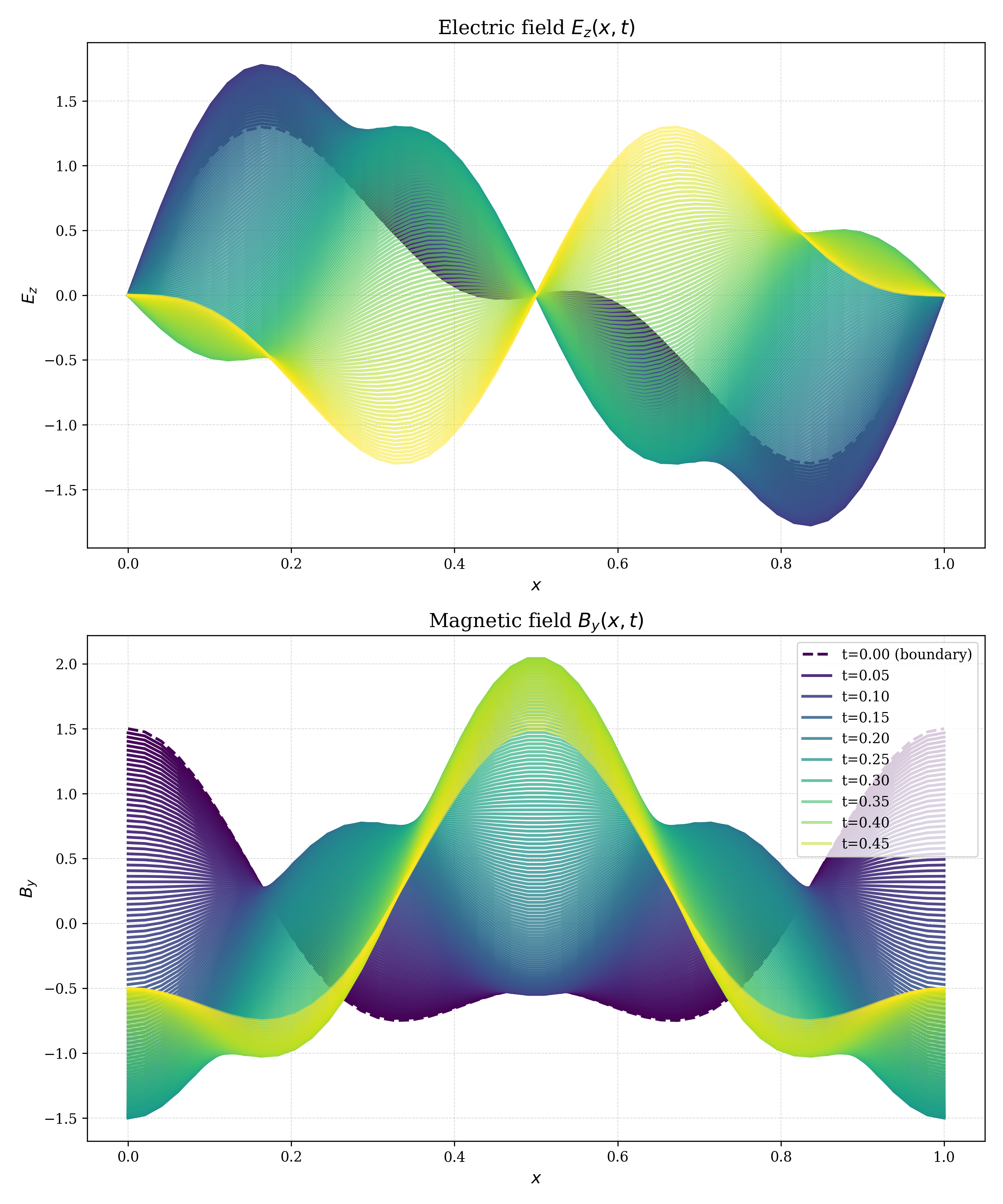}
        \label{fig:maxwell-solution}
    \end{subfigure}
    \hfill
    \begin{subfigure}{0.36\textwidth}
        \centering
        \includegraphics[width=\textwidth]{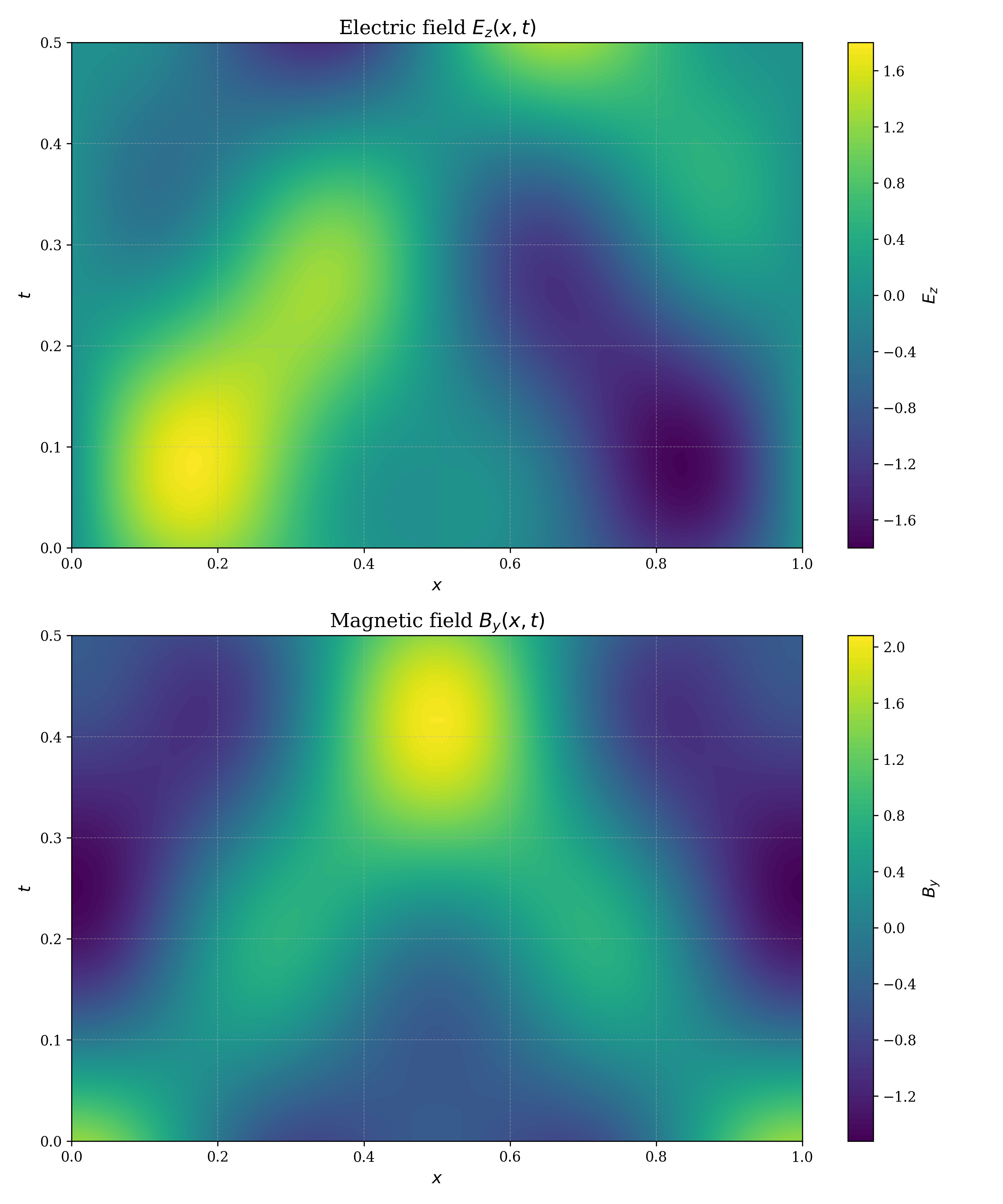}
        \label{fig:maxwell-solution-colormap}
    \end{subfigure}
    \hfill
        \begin{subfigure}{0.25\textwidth}
        \centering
        \includegraphics[width=\textwidth]{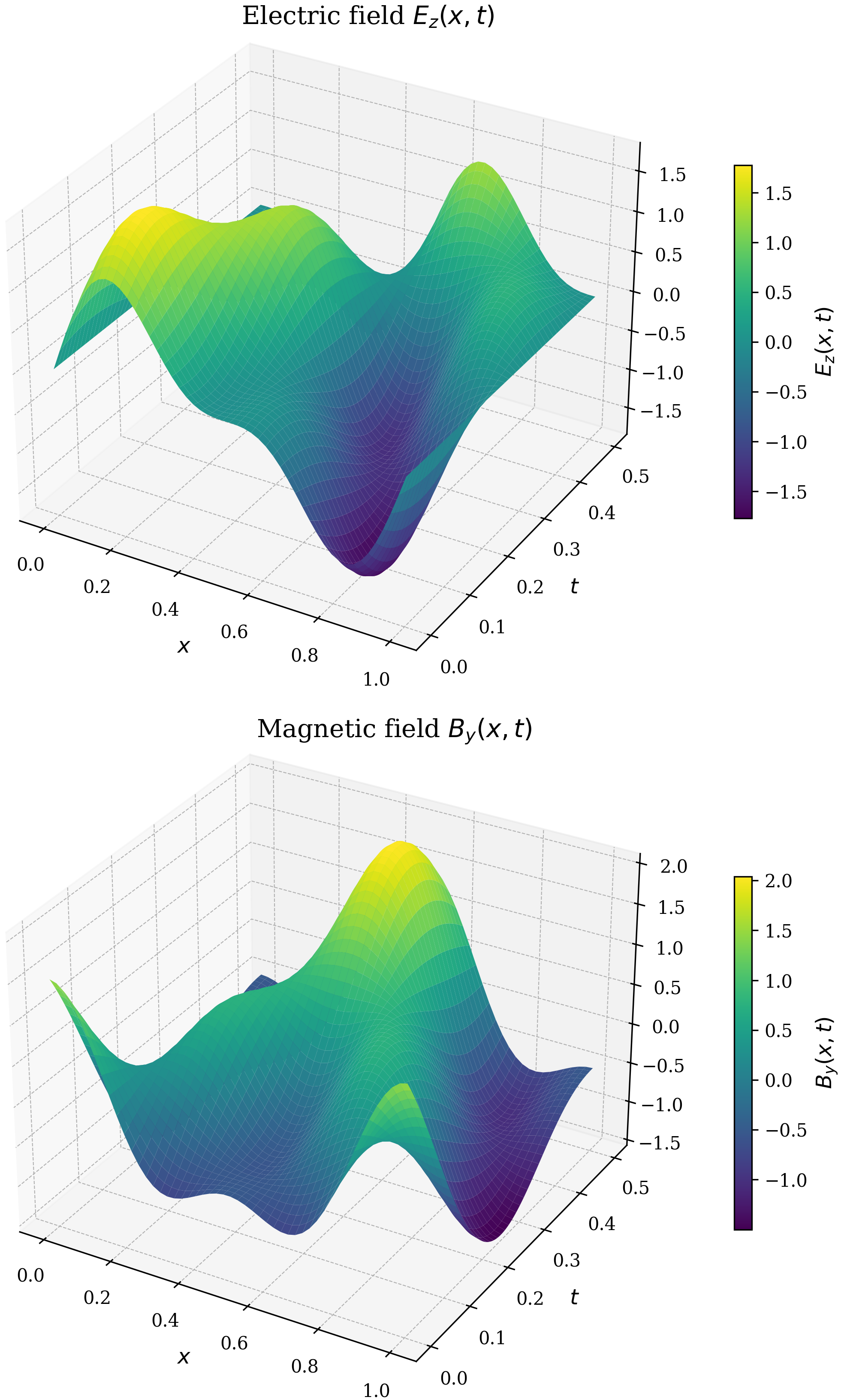}
        \label{fig:maxwell-solution-3d}
    \end{subfigure}
    \caption{Maxwell's equations solution visualization in 1D, 2D and 3D.}
    \label{fig:maxwell-solution-visualization}
\end{figure}

Taking the curl of Faraday's law and Ampère-Maxwell law respectively,
\begin{equation}
\begin{cases}
\nabla \times (\nabla \times \mathbf{E}) = - \frac{\partial}{\partial t} (\nabla \times \mathbf{B}),\\
\nabla \times (\nabla \times \mathbf{B}) = \mu_0 (\nabla \times \mathbf{J}) + \mu_0 \epsilon_0 \frac{\partial}{\partial t} (\nabla \times \mathbf{E}).
\end{cases}
\end{equation}
By vector calculus identity of
\(\nabla \times (\nabla \times \mathbf{E}) = \nabla (\nabla \cdot \mathbf{E}) - \Delta \mathbf{E}\),
and plugging in Ampère-Maxwell law on the right-hand side of the first
equation,
\begin{equation}
\begin{cases}
\nabla (\nabla \cdot \mathbf{E}) - \Delta \mathbf{E} = - \mu_0 \frac{\partial \mathbf{J}}{\partial t} - \mu_0 \epsilon_0 \frac{\partial^2 \mathbf{E}}{\partial t^2}, \\
\nabla (\nabla \cdot \mathbf{B}) - \Delta \mathbf{B} = \mu_0 (\nabla \times \mathbf{J})  + \mu_0 \epsilon_0 \frac{\partial}{\partial t} (\nabla \times \mathbf{E}).
\end{cases}
\end{equation}
By substituting Gauss's law for \(\nabla \cdot \mathbf{E}\) in the first
equation, Gauss's law for magnetism for \(\nabla \cdot \mathbf{B}\) and
Faraday's law for \(\nabla \times \mathbf{E}\) in the second equation,
the two equations after rearrangement are inhomogeneous, i.e.~including
source terms \(\mathbf{F}(\mathbf{r}, t)\), wave equations, taking the
forms of \(c^2 \Delta u - u_{tt} = \mathbf{F}\),
\begin{equation}
\begin{cases}
\Delta \mathbf{E} - \mu_0 \epsilon_0 \frac{\partial^2 \mathbf{E}}{\partial t^2} = \nabla (\frac{\rho}{\epsilon_0}) + \mu_0 \frac{\partial \mathbf{J}}{\partial t}, \\
\Delta \mathbf{B} - \mu_0 \epsilon_0 \frac{\partial^2 \mathbf{B}}{\partial t^2} = - \mu_0 (\nabla \times \mathbf{J}).
\end{cases}
\label{eq:maxwell-wave}
\end{equation}
To simplify the problem, we take the one-dimensional (1D) electromagnetic
wave propagating along \(x\)-axis without sources, i.e.~\(\rho = 0\) and
\(\mathbf{J} = 0\), with the electric field
\(\mathbf{E}(\mathbf{r}, t) = (0, 0, E_z(x, t))\) along \(z\)-axis and
magnetic field \(\mathbf{B}(\mathbf{r}, t) = (0, B_y(x, t), 0)\) along
\(y\)-axis respectively. By expanding the defintion of curl
\(\nabla \times\) operators, and removing the zero terms,
\begin{equation}
\nabla \times \mathbf{E} = \left( \frac{\partial E_z}{\partial y} - \frac{\partial E_y}{\partial z}, \frac{\partial E_x}{\partial z} - \frac{\partial E_z}{\partial x}, \frac{\partial E_y}{\partial x} - \frac{\partial E_x}{\partial y} \right) = \left( 0, - \frac{\partial E_z}{\partial x}, 0 \right),
\end{equation}
thus the reduced last two equations of Maxwell's equations
\eqref{eq:maxwell-differential} are,
\begin{equation}
\begin{cases}
\frac{\partial E_z}{\partial x} = - \frac{\partial B_y}{\partial t}, \\
\frac{\partial B_y}{\partial x} = - \mu_0 \epsilon_0 \frac{\partial E_z}{\partial t}.
\end{cases}
\label{eq:maxwell-1d}
\end{equation}
By taking partial derivatives \(\partial_x\) and \(\partial_t\), and
simplifying, the 1D wave solutions
are\footnote{The 1D wave equation aligned with the general inhomogeneous wave equation \eqref{eq:maxwell-wave}, with $\mathbf{F} = 0$.},
\begin{equation}
\begin{cases}
\frac{\partial^2 E_z}{\partial x^2} - \mu_0 \epsilon_0 \frac{\partial^2 E_z}{\partial t^2} = 0, \\
\frac{\partial^2 B_y}{\partial x^2} - \mu_0 \epsilon_0 \frac{\partial^2 B_y}{\partial t^2} = 0.
\end{cases}
\label{eq:maxwell-1d-wave}
\end{equation}
The initial conditions at \(t=0\) are given as follows,
\begin{equation}
E_z(x, 0) = f(x), \quad B_y(x, 0) = g(x), \quad x \in [x_0, x_f].
\label{eq:maxwell-1d-ic}
\end{equation}
Let \(u = E_z + B_y\) and \(v = E_z - B_y\) and with change of variables
\(x_t = x_{t=0} \pm c t\), where \(c\) defined in
\eqref{eq:mu-epsilon-c} is the speed of light in vacuum. According to
d'Alembert's formula \cite{dalembert1747corde},
\begin{equation}
\begin{cases}
u(x, t) = u(x - c t, 0) = f(x - c t) + g(x - c t), \\
v(x, t) = v(x + c t, 0) = f(x + c t) - g(x + c t). 
\end{cases}
\end{equation}
By reversing the change of variables, the analytical solutions to
\eqref{eq:maxwell-1d-wave} are,
\begin{equation}
\begin{cases}
E_z = \frac{1}{2} ( u + v ) = \frac{1}{2} [f(x - c t) + f(x + c t)] + \frac{1}{2} [g(x - c t) - g(x + c t)], \\
B_y = \frac{1}{2} ( u - v ) = \frac{1}{2} [f(x - c t) - f(x + c t)] + \frac{1}{2} [g(x - c t) + g(x + c t)].
\end{cases}
\label{eq:maxwell-1d-wave-solution}
\end{equation}
The solution is visualized in Figure
\ref{fig:maxwell-solution-visualization}, with initial condition given
as,
\begin{equation}
f(x) = \sin(2 \pi x) + 0.5 \sin(4 \pi x), \; g(x) = \cos(2 \pi x) + 0.5 \cos(4 \pi x) \quad x \in [0, 1], t \in [0, 0.5].
\end{equation}

\hypertarget{viscous-burgers-equation}{%
\subsection{Viscous Burgers' equation}\label{viscous-burgers-equation}}

Viscous Burgers' equation \cite{Takamoto2022PDEBench} captures both non-linear advection, also known as convection
and diffusion phenomena in dynamics,
\begin{equation}
\begin{aligned}
\begin{cases}
\frac{\partial u(x, t)}{\partial t} + u(x, t) \frac{\partial u(x, t)}{\partial x} = \nu \frac{\partial^2 u(x, t)}{\partial x^2}, \quad x \in [x_0, x_f], t \in [t_0, t_f], \\
u(x, 0) = u_0(x), \quad x \in [x_0, x_f], \quad \text{initial condition},
\end{cases}
\end{aligned}
\label{eq:burgers-general}
\end{equation}
where viscosity \(\nu \in \mathbb{R}^+\) is the positive constant, and
\(u_0(x)\) is the initial condition given at \(t=0\). By Cole-Hopf
transformation \cite{Hopf1950BurgersEquation}, unknown function
\(u(x, t)\) is converted into \(\phi(x, t)\) via,
\begin{equation}
u(x, t) = - 2 \nu \frac{\partial}{\partial x} \ln \phi(x, t) = - 2 \nu \frac{1}{\phi(x, t)} \frac{\partial \phi(x, t)}{\partial x} \equiv - 2 \nu \frac{\phi_x}{\phi}.
\label{eq:cole-hopf}
\end{equation}
By chain rule and quotient rule of differentiation, the first-order and
second-order spatial or temporal derivatives of \(u(x, t)\) are,
\begin{equation}
\begin{aligned}
& \frac{\partial u(x, t)}{\partial x} 
= 2 \nu \left( \frac{\phi_x^2}{\phi^2} - \frac{\phi_{xx}}{\phi} \right), \quad 
\frac{\partial u(x, t)}{\partial t} 
= 2 \nu \left(\frac{\phi_x \phi_t}{\phi^2} -  \frac{\phi_{xt}}{\phi}  \right), 
\quad \\
& \frac{\partial^2 u(x, t)}{\partial x^2} 
= 2 \nu \left( \frac{3 \phi_x \phi_{xx}}{\phi^2} - \frac{2 \phi_x^3}{\phi^3} - \frac{\phi_{xxx}}{\phi} \right).
\end{aligned}
\label{eq:burgers-derivatives}
\end{equation}
By plugging \eqref{eq:burgers-derivatives} into
\eqref{eq:burgers-general} and simplifying,
\begin{equation}
2 \nu \left( \frac{\phi_x \phi_t}{\phi^2} - \frac{\phi_{xt}}{\phi} - \nu \frac{\phi_x \phi_{xx}}{\phi^2} + \nu \frac{\phi_{xxx}}{\phi} \right) = 0, 
\quad x \in [x_0, x_f], t \in [t_0, t_f],
\label{eq:burgers-intermediate}
\end{equation}
With the inversion of quotient rule, \eqref{eq:burgers-intermediate} is
rearranged as,
\begin{equation}
2 \nu \frac{\partial}{\partial x} \left( \frac{\nu \phi_{xx}-\phi_t}{\phi} \right) = 0,
\quad x \in [x_0, x_f], t \in [t_0, t_f].
\label{eq:burgers-rearranged}
\end{equation}
By integrating \eqref{eq:burgers-rearranged} with respect to \(x\) and
introducing an integration function \(f(t)\),
\begin{equation}
\frac{\nu \phi_{xx}-\phi_t}{\phi} = f(t),
\quad x \in [x_0, x_f], t \in [t_0, t_f].
\label{eq:burgers-integrated}
\end{equation}
Now introduce \(f(t) = \frac{d F(t)}{dt}\) and
\(\tilde{\phi} = \phi \cdot e^{F(t)}\), thus the derivatives of
\(\tilde{\phi}\) are,
\begin{equation}
\frac{\partial \tilde{\phi}}{\partial t} = e^{F(t)} \left( \phi_t + \phi \frac{d F(t)}{dt} \right), \quad
\frac{\partial^2 \tilde{\phi}}{\partial x^2} = e^{F(t)} \phi_{xx},
\end{equation}
by plugging them into \eqref{eq:burgers-integrated}. The resulting
equation is reduced to the standard heat equation,
\begin{equation}
\nu \frac{\partial^2 \tilde{\phi}(x, t)}{\partial x^2} - \frac{\partial \tilde{\phi}(x, t)}{\partial t} = 0,
\quad x \in [x_0, x_f], t \in [t_0, t_f].
\label{eq:burgers-heat-eq}
\end{equation}
The solution of \eqref{eq:burgers-heat-eq} is formed by heat kernel
\(\Phi(x, t)\) convolved with the initial condition
\(\tilde{\phi}_0 (x) = \tilde{\phi}(x, 0)\) \cite{Evans2010PDE},
\begin{equation}
\tilde{\phi}(x, t) = \int_{-\infty}^{\infty} \Phi(x - x', t) \tilde{\phi}_0 (x') \; dx', \quad \text{where} \quad
\Phi(x, t) = \frac{1}{\sqrt{4 \pi \nu t}} e^{-\frac{x^2}{4 \nu t}}.
\label{eq:burger-heat-equation-solution}
\end{equation}
Note that the transformation from \(\phi\) to \(\tilde{\phi}\) does not
change the Cole-Hopf transformation \eqref{eq:cole-hopf}, since the
additional multiplicative term \(e^{F(t)}\) is independent of \(x\),
\begin{equation}
u(x, t) = - 2 \nu \frac{\partial}{\partial x} \ln \phi(x, t) = - 2 \nu \frac{\partial}{\partial x} \ln \tilde{\phi}(x, t).
\label{eq:cole-hopf-tilde}
\end{equation}
From the Cole-Hopf \eqref{eq:cole-hopf-tilde} at \(t=0\) and via
integration, the initial condition for \(\tilde{\phi}(x, 0)\) is thus,
\begin{equation}
\tilde{\phi}_0 (x) = - \frac{1}{2 \nu} \int_{0}^{x} u_0 (x') \; dx'.
\label{eq:burger-heat-initial}
\end{equation}
The analytical solution of \eqref{eq:burgers-general} is thus plugging
\eqref{eq:burger-heat-initial} into
\eqref{eq:burger-heat-equation-solution} and then into
\eqref{eq:cole-hopf-tilde}.

\begin{figure}[htpb]
    \centering
    \begin{subfigure}{0.36\textwidth}
        \centering
        \includegraphics[width=\textwidth]{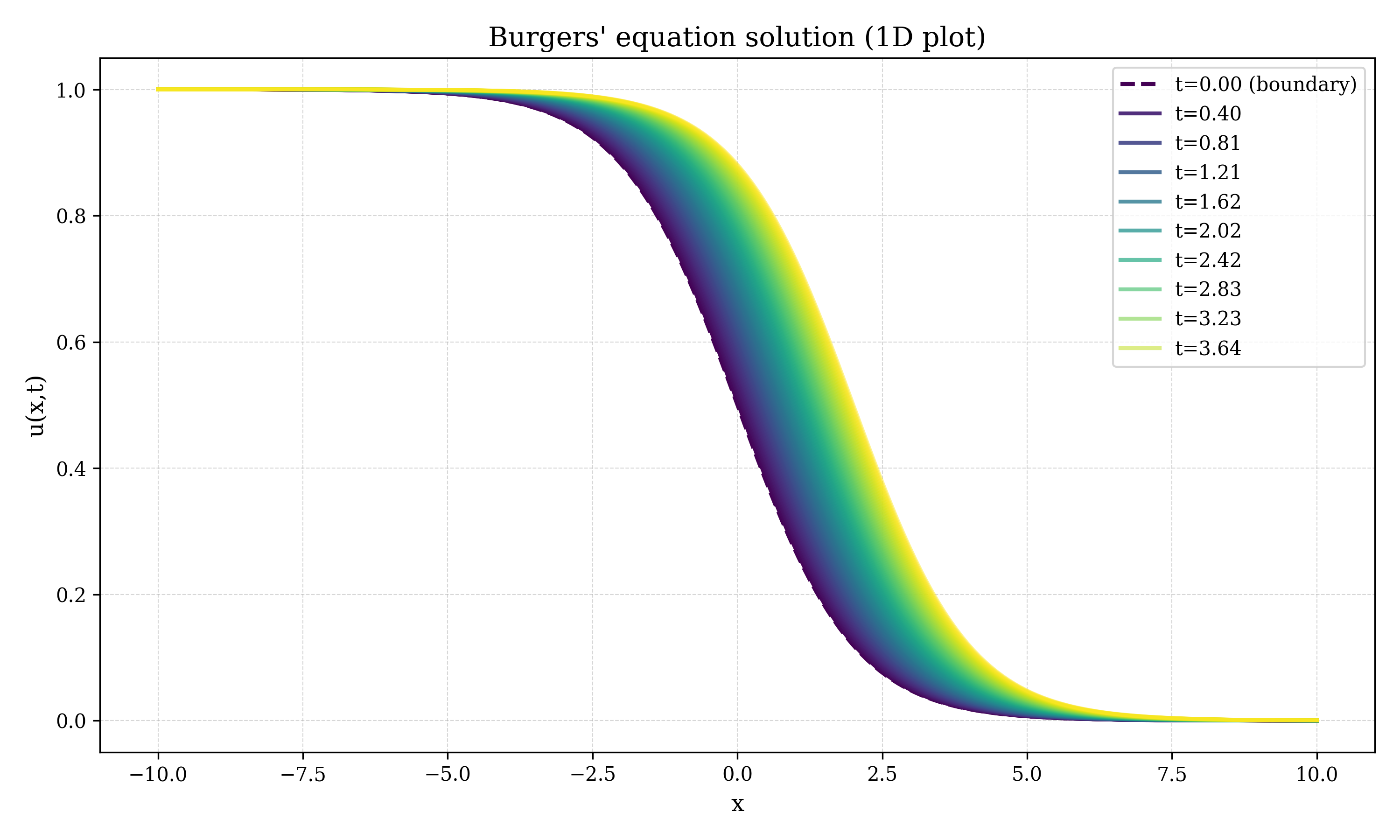}
        \label{fig:burgers-solution}
    \end{subfigure}
    \hfill
    \begin{subfigure}{0.36\textwidth}
        \centering
        \includegraphics[width=\textwidth]{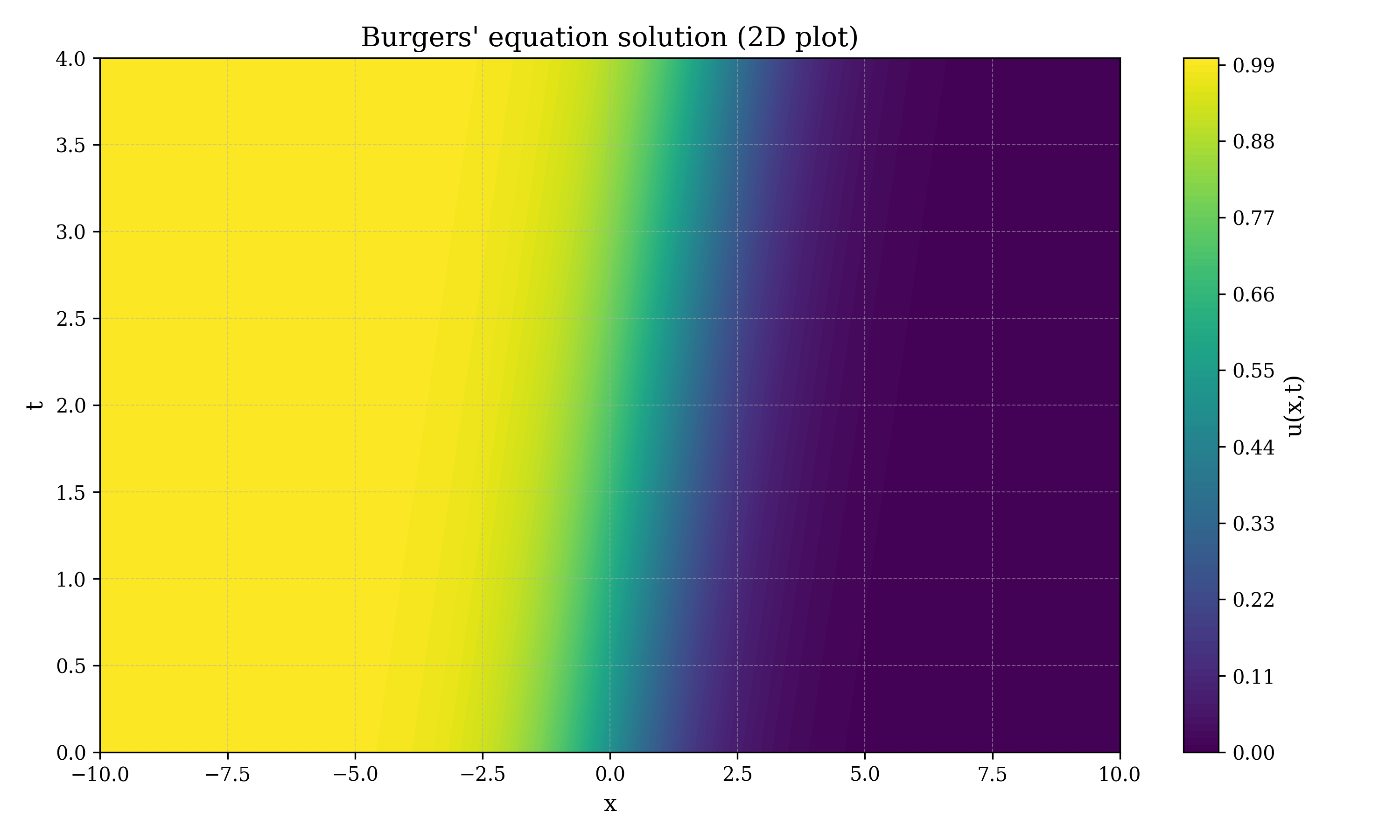}
        \label{fig:burgers-solution-colormap}
    \end{subfigure}
    \hfill
        \begin{subfigure}{0.24\textwidth}
        \centering
        \includegraphics[width=\textwidth]{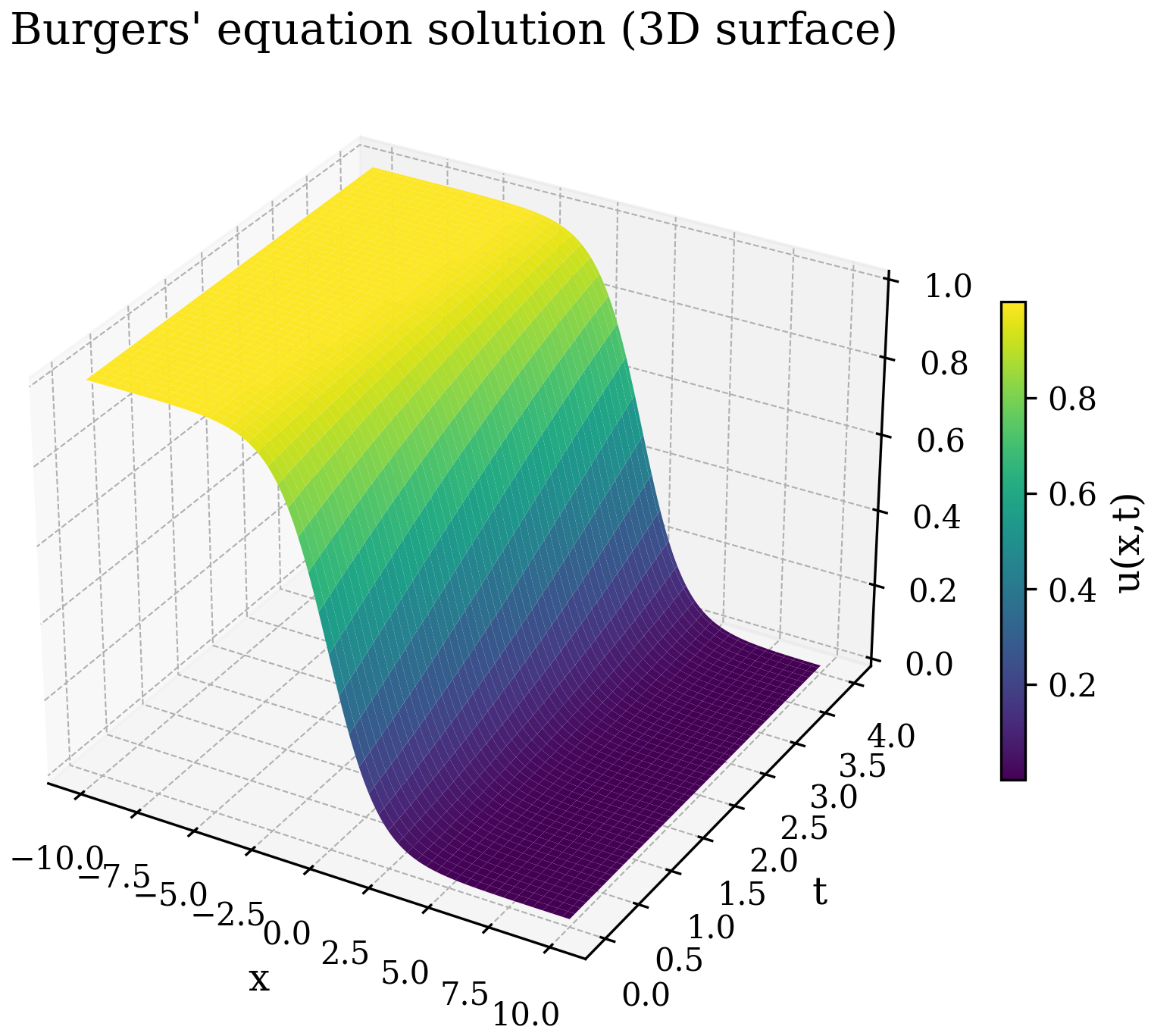}
        \label{fig:burgers-solution-3d}
    \end{subfigure}
    \caption{Burgers' equation solution visualization in 1D, 2D and 3D.}
    \label{fig:burgers-solution-visualization}
\end{figure}

In this work, we consider when \(u_0(- \infty)\) and \(u_0(\infty)\)
exist and \(u_0'(x) < 0\) for all \(x \in \mathbb{R}\), the explicit
expression \cite{Bateman1915FluidMotion} is then a steadily propagating
wave as below,
\begin{equation}
u(x, t) = c - \Delta u_0 \tanh \left( \frac{\Delta u_0}{2 \nu} (x - c t) \right), \quad \text{where } c = \frac{u_0(- \infty) + u_0(\infty)}{2}, \Delta u_0 = \frac{u_0(- \infty) - u_0(\infty)}{2}.
\label{eq:burgers-sol-steady}
\end{equation}
The solution is visualized in Figure
\ref{fig:burgers-solution-visualization}, with initial condition set by
\eqref{eq:burgers-sol-steady}, \(u_0(- \infty) = 1\),
\(u_0(\infty) = 0\), and \(\nu = 0.5\).

\hypertarget{pde-solvers}{%
\subsection{PDE solvers}\label{pde-solvers-appendix}}

There are many attempts to solve the PDE solution field \(u\). Among
which, we categorize PDE solvers into two types, i.e.~numerical analysis
methods and neural-based methods.

\textbf{Constrained optimization.}  In PDE solvers, constraints are boundary conditions, initial conditions, or PDE residuals. They can be in the form of either soft or hard constraints. The former is the cost functions that are penalised, while the latter is that can not be violated, e.g.~(in)equality forms.

\begin{longtable}[]{@{}
  >{\raggedright\arraybackslash}p{(\columnwidth - 8\tabcolsep) * \real{0.2000}}
  >{\centering\arraybackslash}p{(\columnwidth - 8\tabcolsep) * \real{0.2857}}
  >{\centering\arraybackslash}p{(\columnwidth - 8\tabcolsep) * \real{0.1714}}
  >{\centering\arraybackslash}p{(\columnwidth - 8\tabcolsep) * \real{0.1714}}
  >{\centering\arraybackslash}p{(\columnwidth - 8\tabcolsep) * \real{0.1714}}@{}}
\caption{Summary of different PDE solvers.}\tabularnewline
\toprule\noalign{}
\begin{minipage}[b]{\linewidth}\raggedright
\textbf{method}
\end{minipage} & \begin{minipage}[b]{\linewidth}\centering
\textbf{motivation}
\end{minipage} & \begin{minipage}[b]{\linewidth}\centering
\textbf{training}
\end{minipage} & \begin{minipage}[b]{\linewidth}\centering
\textbf{supervised}
\end{minipage} & \begin{minipage}[b]{\linewidth}\centering
\textbf{constraint}
\end{minipage} \\
\midrule\noalign{}
\endfirsthead
\toprule\noalign{}
\begin{minipage}[b]{\linewidth}\raggedright
\textbf{method}
\end{minipage} & \begin{minipage}[b]{\linewidth}\centering
\textbf{motivation}
\end{minipage} & \begin{minipage}[b]{\linewidth}\centering
\textbf{training}
\end{minipage} & \begin{minipage}[b]{\linewidth}\centering
\textbf{supervised}
\end{minipage} & \begin{minipage}[b]{\linewidth}\centering
\textbf{constraint}
\end{minipage} \\
\midrule\noalign{}
\endhead
\bottomrule\noalign{}
\endlastfoot
FDM & grid-based & \(\times\) & N/A & hard \\
PINN & physics-driven & \(\checkmark\) & \(\times\) & soft \\
NO & data-driven & \(\checkmark\) & \(\checkmark\) & soft \\
- FNO & data-driven & \(\checkmark\) & \(\checkmark\) & soft \\
- PINO & hybrid & \(\checkmark\) & \(\checkmark\) & soft \\
KM & mesh-free grid & \(\times\) & N/A & hard \\
- CNF & KM on neural fields & \(\checkmark\) & \(\times\) & hard \\
\end{longtable}

\hypertarget{finite-difference-method-fdm}{%
\subsubsection{Finite difference method
(FDM)}\label{finite-difference-method-fdm}}

Considering a discretized sequence
\(\mathbf{u} \in \mathbb{R}^{N\times M}\) of the continuous function
\(u(x, t)\) as in \eqref{eq:advection-general}. Along the spatial
dimension \(x\) and temporal dimension \(t\), there are \(N\) and \(M\)
sampled points respectively. The finite difference operators
\cite{Iserles2008NumericalAnalysis} defined on per element, are as
follows:
\begin{equation}
(\Delta \mathbf{u})_i = \begin{cases}
\begin{aligned}
(\Delta^{+} \mathbf{u})_i &= u_{i+1}^{j} - u_i^j, & \quad \text{forward difference.} \\
(\Delta^{-} \mathbf{u})_i &= u_i^j - u_{i-1}^{j}, & \quad \text{backward difference.} \\
(\Delta^{0} \mathbf{u})_i &= \frac{u_{i+1}^{j} - u_{i-1}^{j}}{2}, & \quad \text{central difference.} 
\end{aligned}
\end{cases},
\label{eq:difference-operators}
\end{equation}
for which \(i \in \{0, 1, \ldots, N-1\}\) is the \emph{spatial index},
and \(j \in \{0, 1, \ldots, M-1\}\) is the \emph{temporal index} of the
sequence \(\mathbf{u}\).

The partial equations often involve full and/or partial derivatives,
where differential operators can be discretized into difference
operators \eqref{eq:difference-operators} via finite difference method
\cite{Bargteil2018}. By Taylor expansion of \(u(x\pm \Delta x, t)\)
around \(u(x, t)\) up to the first order error, the corresponding
examples for the spatial derivative are,
\begin{equation}
\begin{aligned}
\frac{\partial u_i^j(x, t)}{\partial x} 
&= 
\begin{cases}
\frac{(\Delta^{-} \mathbf{u})_i}{\Delta x}  + O(\Delta x)  \approx \frac{(\Delta^{-} \mathbf{u})_i}{\Delta x} = \frac{u_i^j - u_{i-1}^{j}}{\Delta x} & \quad \text{if } \beta > 0, \\
\frac{(\Delta^{+} \mathbf{u})_i}{\Delta x}  + O(\Delta x) \approx \frac{(\Delta^{+} \mathbf{u})_i}{\Delta x}  = \frac{u_{i+1}^{j} - u_i^j}{\Delta x} & \quad \text{if } \beta < 0. \\
\end{cases}, & \quad \text{upwind scheme.} \\
&= \frac{(\Delta^{0} \mathbf{u})_i}{\Delta x}  + O(\Delta x^2) \approx \frac{(\Delta^{0} \mathbf{u})_i}{\Delta x} = \frac{u_{i+1}^{j} - u_{i-1}^{j}}{2 \Delta x}, & \quad \text{central difference.} 
\end{aligned}
\end{equation}
where \(\Delta x\) is the spatial spacing, with spatial index \(i\) and
temporal index \(j\) defined above. The \textbf{upwind scheme}
\cite{Patankar1980NumericalHeat} considers where the information comes
from, e.g.~when \(\beta > 0\), the wave propagates rightwards, and thus
\(u_i^j\) is influenced by \(u_{i-1}^j\), and vice versa for downwind
scheme. Under the upwind scheme, the advection equation
\eqref{eq:advection-general} is therefore as the following ODE,
\begin{equation}
\frac{\partial u(x, t)}{\partial t} + \beta [ \mathbb{I}_{\beta > 0} \frac{(\Delta^{-} \mathbf{u})_i}{\Delta x} + \mathbb{I}_{\beta < 0} \frac{(\Delta^{+} \mathbf{u})_i}{\Delta x} ] = 0,
\label{eq:advection-ode}
\end{equation}
where the indicator function
\(\mathbb{I}_{\beta > 0}=\begin{cases} 1 & \text{if } \beta > 0, \\ 0 & \text{otherwise.} \end{cases}\)
is for controlling different cases of
\(\beta\)\footnote{Alternatively, one may use $\max(\beta, 0)$ and $\min(\beta, 0)$ to replace $\beta \mathbb{I}_{\beta > 0}$ and $\beta \mathbb{I}_{\beta < 0}$ respectively.}.

By forward Euler method for ODEs, the temporal derivative is discretized
via the forward difference operator. After which, the advection equation
\eqref{eq:advection-general} is simplified as, with \(\Delta t\) being
the temporal spacing,
\begin{equation}
\frac{u_i^{j+1} - u_i^j}{\Delta t} + \beta [ \mathbb{I}_{\beta > 0} \frac{(\Delta^{-} \mathbf{u})_i}{\Delta x} + \mathbb{I}_{\beta < 0} \frac{(\Delta^{+} \mathbf{u})_i}{\Delta x} ] = 0.
\end{equation}
With algebraic reordering, the upwind scheme update rule is thus,
\begin{equation}
u_i^{j+1} = u_i^j - \frac{\beta \Delta t}{\Delta x} [ \mathbb{I}_{\beta > 0} (\Delta^{-} \mathbf{u})_i + \mathbb{I}_{\beta < 0} (\Delta^{+} \mathbf{u})_i ].
\label{eq:advection-update}
\end{equation}
\textbf{Stability condition.} For implicit numerical schemes, e.g.~the
backward Euler method, the solution is unconditionally stable. However,
for explicit numerical schemes, e.g.~the forward Euler method above,
stability conditions must be satisfied to avoid numerical instability,
which we briefly introduce below.

In the 1D space, the scalar Courant number \(C\), also known as the CFL
stability criteria, measures the ratio of how far the wave propagates in
one time interval \(\Delta t\) to the spatial spacing \(\Delta x\). The
CFL condition \cite{CFL1928} states that \(C\) must satisfy,
\begin{equation}
C = \frac{|\beta| \Delta t}{\Delta x} \leq C_{max},
\end{equation}
where \(C_{max}\) is a problem-dependent constant. It sets the maximum
allowable time step \(\Delta t\) for a given \(\Delta x\), for numerical
stability.

\hypertarget{physics-informed-neural-network-pinn}{%
\subsubsection{Physics-informed neural network
(PINN)}\label{physics-informed-neural-network-pinn}}

Physics-informed neural network (PINN) \cite{RAISSI2019PINNs} is a
data-driven approach for functional PDE approximation, which requires a
large labeled dataset but has the ability to generalize. Consider the
general form of PDEs defined in \eqref{eq:pde-general}, PINNs
approximate the unknown solution \(u(x, t) \in \mathcal{U}\) with a
neural network \(\hat{u}_{\theta}(x, t) \in \mathcal{U}\),
i.e.~\(\hat{u}_{\theta}(x, t) \approx u(x, t)\), parameterized by
updatable parameters \(\theta \in \Theta\).

\textbf{Residual} \(\mathcal{R}_\theta\) of the PDEs is calculated
without supervised data for the neural network \(\hat{u}_{\theta}\),
which is minimized via automatic differentiation
\cite{Baydin2018AutoDifferentiation}\footnote{Example of $\frac{\partial u}{\partial x}$: \texttt{u\_x = torch.autograd.grad(outputs=u, inputs=x, create\_graph=True)[0]}}
during training for generalizability,
\begin{equation}
\mathcal{R}_\theta (x, t) \in \mathcal{Y} = \mathcal{D}[\hat{u}_{\theta}](x, t) - f(x, t), \quad x \in \Omega, t \in [t_0, t_f].
\label{eq:pinn-residual}
\end{equation}
The residual loss
\(L_R\)\footnote{Note that $L_R$ is the same as the risk $\mathcal{R}$ defined in \eqref{eq:function-risk}, but for the residual $\mathcal{R}_\theta$ instead of the solution $u$.},
also known as the physics-informed loss, is defined to be the \(p\)-norm
of the residual \(\mathcal{R}_\theta\) in \eqref{eq:pinn-residual}.
During training, \(N_\mathcal{R}\) quadrature points are sampled, where
the integral loss is approximated by the discretized loss
\(\mathcal{L}_R\) with weights \(\omega_k\) at each sample index \(k\)
and training error \(\mathcal{E}_{T}(\theta)\),
\begin{equation}
\begin{aligned}
L_R:=
(||\mathcal{R}_\theta ||_p)^p
:= &
\underbrace{
    [(\int_{\mathbb{D}} |\mathcal{R}_\theta|^{p} \; dx \; dt)^{\frac{1}{p}}]^{p} 
}_{\text{integral } L_R}
= \int_{\mathbb{D}} |\mathcal{R}_\theta|^{p} \; dx \; dt \\
\text{By quadrature,} \quad
= & \underbrace{\sum_{k=1}^{N_\mathcal{R}} \omega_{k} | \mathcal{R}_\theta (x_k, t_k) |^p}_{\text{discretized } \mathcal{L}_R} + \mathcal{E}_{T}(\theta)
\approx  
\mathcal{L}_R, 
\quad \text{where } 
\mathcal{E}_{T}(\theta) = L_R - \mathcal{L}_R.
\end{aligned}
\label{eq:pinn-training-error}
\end{equation}
If considering the boundary conditions, the residual for the \(i\)-th
boundary condition \(\mathcal{R}^{\mathcal{B}_i}_\theta\) is calculated
via \eqref{eq:pde-general} as well, after which the boundary condition
loss \(\mathcal{L}_{\text{BC}}\) is defined accordingly,
\begin{equation}
\mathcal{R}^{\mathcal{B}_i}_\theta (x, t) \in \mathcal{Z}_i = \mathcal{B}_i[\hat{u}_{\theta}](x, t) - g_i(x, t), \quad x \in \partial \Omega_i, t \in [t_0, t_f].
\label{eq:pinn-bc-residual}
\end{equation}
As defined in \eqref{eq:error-decomposition}, the \emph{total error}
between the optimal solution from the network \(\hat{u}_{\theta}\) and
the ground truth \(u\) is, by expanding \eqref{eq:function-risk},
\begin{equation}
\mathcal{E}_{\text{PINN}}(\theta) = (||\hat{u}_{\theta} - u||_p)^{p}.
\label{eq:pinn-total-error}
\end{equation}
During training, the network is optimized on supervised dataset
\(\{(x_n, t_n), u(x_n, t_n)\}_{n=1}^{N_d}\), with \(N_d\) being the
total number of data. The supervised loss
\(\mathcal{L}_{\text{data}}\)\footnote{Note that when the supervised data is only sampled on the boundary, the supervised loss and the boundary condition loss are the same.}
approximates the total error \eqref{eq:pinn-total-error},
\begin{equation}
\mathcal{L}_{\text{data}} = \frac{1}{N_d} \sum_{n=1}^{N_d} (|\hat{u}_{\theta}(x_n, t_n) - u(x_n, t_n)|^p).
\label{eq:pinn-supervised-loss}
\end{equation}
\textbf{Training}. PINN approximates the solution as
\(\hat{u}_{\theta} = u_{\theta^{\mathrm{opt}}}(x, t)\). To avoid
overfitting due to the \emph{limited} supervised data, the main goal is
to minimize the unsupervised residual error \(\mathcal{L_R}\)
\eqref{eq:pinn-training-error}. With the addition of the supervised loss
\eqref{eq:pinn-supervised-loss} and the boundary condition residual
\eqref{eq:pinn-bc-residual}, the optimized theta is
\(\theta^{\mathrm{opt}} \approx \arg \min_{\theta \in \Theta} \mathcal{L}\),
where the total training loss \(\mathcal{L}_{\text{PINN}}\) is,
\begin{equation}
\mathcal{L}_{\text{PINN}} = \underbrace{ \sum_{k=1}^{N_\mathcal{R}} \omega_{k} | \mathcal{R}_\theta (x_k, t_k) |^p }_{\text{Discretized residual loss } \mathcal{L_R}} + 
\lambda_1
\underbrace{\frac{1}{N_d} \sum_{n=1}^{N_d} (|\hat{u}_{\theta}(x_n, t_n) - u(x_n, t_n)|^p)}_{\text{Supervised loss } \mathcal{L}_{\text{data}}} +
\lambda_2
\underbrace{\sum_{i} \sum_{b=1}^{N_{\mathcal{B}_i}} \omega_b^{\mathcal{B}_i} | \mathcal{R}^{\mathcal{B}_i}_\theta (x_b, t_b) |^p}_{\text{BC loss } \mathcal{L}_{\text{BC}}},
\label{eq:pinn-training-loss-sum}
\end{equation}
with weights \(\omega_b^{\mathcal{B}_i}\) at each sample index \(b\) for
\(i\)-th boundary condition and regularization parameters
\(\lambda_1, \lambda_2 > 0\) for combining different losses.

\begin{algorithm}
\caption{Physics-Informed Neural Network training pseudocode.}
\label{alg:pinn-training}
\begin{algorithmic}[1]
\State \textbf{Input:} Initial parameters $\theta$ for network $\hat{u}_{\theta}$.
\State \textbf{Output:} Optimized parameters $\theta^{\mathrm{opt}}$ for network $\hat{u}_{\theta}$.
\State \textbf{Hyperparameters:} Learning rate $\eta$, number of training iterations $N_{\text{iter}}$.
\While{number of iterations $< N_{\text{iter}}$}
    \State Sample PDE points $x_k \in \Omega, t_k \in [t_0, t_f]$ and boundary points $x_b \in \partial \Omega_i, t_b \in [t_0, t_f]$.
    \State Compute the network outputs$\hat{u}_{\theta}$ and their derivatives $\mathcal{D}[\hat{u}_{\theta}]$ and boundary $\mathcal{B}_i[\hat{u}_{\theta}]$.
    \State Compute loss $\mathcal{L}_{\text{PINN}} = \mathcal{L_R} + \mathcal{L}_{\text{data}} + \mathcal{L}_{\text{BC}}$ by \eqref{eq:pinn-training-loss-sum}.
    \State By gradient descent, update $\theta \leftarrow \theta - \eta \, \nabla_\theta \mathcal{L}$.
\EndWhile
\end{algorithmic}
\end{algorithm}


\hypertarget{neural-operator-no}{%
\subsubsection{Neural operator (NO)}\label{neural-operator-no}}

\textbf{Operator learning}. From the general form of PDEs
\eqref{eq:pde-general}, we assume that within the \(\mathcal{D}\)
operator, the source function \(f\) or the initial conditions, there is
a parameter \(a\) of the same dimension as the solution \(u\). In this
subsection, we denote the differential operator \(\mathcal{D}\) as
\(\mathcal{D}_a\), where the PDE is thus \(\mathcal{D}_a[u] = f\).

Given the dataset
\(\{(a_i^j, f_i^j), u_i^j | i= 1, ..., N_\mathcal{R} \}_{j=1}^{N_{pde}}\)
with \(N_{pde}\) PDE instances each \(N_\mathcal{R}\) quadrature points,
the idea of operator learning \cite{Li2020FNO} is to learn the operator
\(\mathcal{G}\) mapping input
\(a \in \mathcal{A}: \mathbb{D} \to \mathbb{R}\) to the solution
\(u \in \mathcal{U}: \mathbb{D} \to \mathbb{R}\),
i.e.~\(\mathcal{G} (a, f) = u\) connecting two function spaces
\(\mathcal{A}\) and \(\mathcal{U}\) with \emph{infinite} dimensions,
which is \textbf{challenging} for neural networks since they are for
\emph{finite} dimensions instead.

\textbf{Solution A.} To solve this challenge, one solution is to
\textbf{parameterize} the PDE \(u = u(t, x, \mu)\), assuming that \(a\)
is measureable in finite dimension,
i.e.~\(a = a(\mu), \mu \in \mathbb{R}^{d_y}\). It's a techinique widely
used in aircraft design and manufacturing \cite{Zhang2009PDEAircraft},
image processing \cite{Perez2003PoissonImageEditing}. The
\textbf{training} process is therefore to minimize the supervised loss
\(\mathcal{L}_{data}\) from data with \(p\)-norm, which measures how
much the predicted solution \(\mathcal{G}_{\theta}(a_i, f_i)\) deviates
from the ground truth \(u_i\) for each data point \(i\),
\begin{equation}
\mathcal{L}_{data} = \frac{1}{N_{pde} \times N_\mathcal{R}} \sum_{j=1}^{N_{pde}} \sum_{i=1}^{N_{\mathcal{R}}}  (||\mathcal{G}_{\theta}(a_i^j, f_i^j) - u_i^j||_p)^p.
\end{equation}
The approximated solution is therefore
\(\hat{u}_{\theta} = \mathcal{G}_{\theta^{\mathrm{opt}}}(a, f)\), where
\(\theta^{\mathrm{opt}} \approx \arg \min_{\theta \in \Theta} \mathcal{L}_{data}\).
There is no addition of the PDE residual loss \(\mathcal{L_R}\)
\eqref{eq:pinn-residual} as in PINNs for basic operator learning. An
extension, termed as \textbf{physics-informed neural operator (PINO)}
\cite{Li2024PINOs}, combines the supervised loss \(\mathcal{L}_{data}\)
and the residual loss \(\mathcal{L_R}\) as the total training loss,
\begin{equation}
\mathcal{L}_{\text{PINO}} = \mathcal{L}_{data} + \lambda \underbrace{\frac{1}{N_{pde}  \times N_\mathcal{R}} \sum_{j=1}^{N_{pde}} \sum_{i=1}^{N_\mathcal{R}} (||\mathcal{D}_a[\mathcal{G}_{\theta}(a_i^j, f_i^j)] - f_i^j(x_i, t_i)||_p)^p}_{\text{Residual loss } \mathcal{L_R}},
\end{equation}



Despite the simplicity in ideas, the parameterization \emph{suffers
from} how to sample from the given space, non-uniqueness, and low
generalization to unseen \(a\).

\textbf{Solution B.} Interpolation from the discretized grid, including
a neural network based interpolator or traditional methods (linear,
cubic, spline, etc). However, it \emph{suffers from} inconsistency
between the discretized and continuous functions.


\textbf{Solution C.} Generalize the neural network from discrete to
continuous function space.
\begin{equation}
(\mathcal{N}_l v)(x) = \sigma [A_l v(x) + B_l(x) + \int_D K_l(x, y) v(y) \; dy], \quad x \in D.
\end{equation}
Fast implemenentation via FFT.

\section{Architecture details} \label{arch-details}
We adopt a multi-scale feed-forward neural network architecture for PINN \cite{Wang2021MutliScalePINNs}, which augments the original feed-forward architecture with input encoding layers of multiple frequency scales. There are three hidden layers each with 64 neurons. For FNO, we adopt the architecture with 16 modes retained in the spectral convolution layer, and the latent feature dimension is 64. 

\hypertarget{environment-setup}{%
\section{Environment setup}\label{environment-setup}}

All the measurements are conducted on a Mac M1, with a single-core CPU
running at 3.2 GHz. In the following sections, the data points are
uniformly sampled unless otherwise specified.

\begin{longtable}[]{@{}
  >{\centering\arraybackslash}p{(\columnwidth - 6\tabcolsep) * \real{0.1639}}
  >{\centering\arraybackslash}p{(\columnwidth - 6\tabcolsep) * \real{0.1639}}
  >{\centering\arraybackslash}p{(\columnwidth - 6\tabcolsep) * \real{0.1639}}
  >{\centering\arraybackslash}p{(\columnwidth - 6\tabcolsep) * \real{0.5082}}@{}}
\caption{\label{tab:maxwell-1d-setup} 1D Maxwell's equations
experimental setup.}\tabularnewline
\toprule\noalign{}
\begin{minipage}[b]{\linewidth}\centering
\textbf{domain}
\end{minipage} & \begin{minipage}[b]{\linewidth}\centering
\textbf{time span}
\end{minipage} & \begin{minipage}[b]{\linewidth}\centering
\textbf{parameter}
\end{minipage} & \begin{minipage}[b]{\linewidth}\centering
\textbf{initial conditions}
\end{minipage} \\
\midrule\noalign{}
\endfirsthead
\toprule\noalign{}
\begin{minipage}[b]{\linewidth}\centering
\textbf{domain}
\end{minipage} & \begin{minipage}[b]{\linewidth}\centering
\textbf{time span}
\end{minipage} & \begin{minipage}[b]{\linewidth}\centering
\textbf{parameter}
\end{minipage} & \begin{minipage}[b]{\linewidth}\centering
\textbf{initial conditions}
\end{minipage} \\
\midrule\noalign{}
\endhead
\bottomrule\noalign{}
\endlastfoot
\(x_0=0,\) \(x_f=1\) & \(t_0=0,\) \(t_f=\frac{1}{2}\) & \(c=1\) &
\(E_z(x, 0) = \sin(2 \pi x) + \frac{1}{2}\sin(4 \pi x)\)
\(B_y(x, 0) = \cos(2 \pi x) + \frac{1}{2}\cos(4 \pi x)\) \\
\end{longtable}

\section{Burgers' equation stability experiment}

\textbf{Stability}. For four time discretization schemes, only forward Euler scheme is unstable (Table
\ref{tab:Stability-forward-euler}), where time step 
\(\Delta t\) exceeds the stability limit when \(C^t_{\text{scale}} = 1\)
and \(2\) according to CFL condition.

\begin{longtable}[]{@{}
  >{\centering\arraybackslash}p{(\columnwidth - 8\tabcolsep) * \real{0.2000}}
  >{\centering\arraybackslash}p{(\columnwidth - 8\tabcolsep) * \real{0.2000}}
  >{\centering\arraybackslash}p{(\columnwidth - 8\tabcolsep) * \real{0.2000}}
  >{\centering\arraybackslash}p{(\columnwidth - 8\tabcolsep) * \real{0.2000}}
  >{\centering\arraybackslash}p{(\columnwidth - 8\tabcolsep) * \real{0.2000}}@{}}
\caption{\label{tab:Stability-forward-euler} Stability test of forward
Euler Kansa method on Burgers' equation.}\tabularnewline
\toprule\noalign{}
\begin{minipage}[b]{\linewidth}\centering
\(C^t_{\text{scale}}\)
\end{minipage} & \begin{minipage}[b]{\linewidth}\centering
1
\end{minipage} & \begin{minipage}[b]{\linewidth}\centering
2
\end{minipage} & \begin{minipage}[b]{\linewidth}\centering
4
\end{minipage} & \begin{minipage}[b]{\linewidth}\centering
10
\end{minipage} \\
\midrule\noalign{}
\endfirsthead
\toprule\noalign{}
\begin{minipage}[b]{\linewidth}\centering
\(C^t_{\text{scale}}\)
\end{minipage} & \begin{minipage}[b]{\linewidth}\centering
1
\end{minipage} & \begin{minipage}[b]{\linewidth}\centering
2
\end{minipage} & \begin{minipage}[b]{\linewidth}\centering
4
\end{minipage} & \begin{minipage}[b]{\linewidth}\centering
10
\end{minipage} \\
\midrule\noalign{}
\endhead
\bottomrule\noalign{}
\endlastfoot
\(\mathcal{\hat{R}}_{\text{relative } L_2}\) & \(3.74 \times 10^{29}\) &
NaN & \(4.31 \times 10^{-3}\) & \(3.11 \times 10^{-3}\) \\
Stability & \emph{unstable} & \emph{unstable} & \textbf{stable} &
\textbf{stable} \\
\end{longtable}

\section{Learning theory}  \label{learning-theory}
\hypertarget{functional-analysis}{%
\subsection{Functional analysis}\label{functional-analysis}}

We introduce basic functional analysis concepts here for PDE solvers and
later learning theory (Appendix § \ref{learning-theory}).

The \(p\)-\textbf{norm} of a function
\(f: \Omega \subseteq \mathbb{R}^d \to \mathbb{R}\) is defined as,
\begin{equation}
||f||_p = ( \int_{\Omega} |f(x)|^p \; dx )^{\frac{1}{p}}, \text{for } 1 \leq p < \infty.
\label{eq:p-norm}
\end{equation}
The \(p\)-\textbf{integrable} function \(f \in L^p(\Omega)\), is defined
as
\begin{equation}
(||f||_p)^p =
\int_{\Omega} |f(x)|^p \; dx < \infty.
\label{eq:p-integrable}
\end{equation}
For additional concepts used for learning theory, please refer to
Appendix § \ref{functional-analysis-addendum}.

\hypertarget{functional-analysis-addendum}{%
\subsection{Functional analysis
addendum}\label{functional-analysis-addendum}}

\textbf{Smoothness} \cite{Rudin1976MathAnalysis} of a function is
defined to be the number of continuous derivatives it has. The class of
function \(f\) with smoothness \(k \in \mathbb{N}^+\) has at least a
\(k\)-th derivative, and is denoted as \(f \in C^k\),
\begin{equation}
C^k(\Omega) = \{ f: \Omega \subseteq \mathbb{R}^d \to \mathbb{R} \; | \; 
\forall \alpha \leq k. \;
\partial^{\alpha} f \text{ exists and is continuous}  \}.
\end{equation}
When \(k = \infty\), the function is differentiable at all orders. While
not every function is not smooth, there is a generalization of smooth
functions, i.e.~Sobolev functions.

The \textbf{weak derivative} \(f'\) generalizes to include functions
that are not differentiable, but locally integrable on bounded domain
\([a, b]\). The \(f'\) definition is for all smooth test functions
\(\phi\), with \(\phi(a) = \phi(b) = 0\),
\begin{equation}
\begin{aligned}
\int_{a}^{b} f(x) \phi'(x) \; dx &= [f(x) \phi(x)]_{a}^{b} - \int_{a}^{b} f'(x) \phi(x) \; dx, \quad \text{by integration by parts}, \\
&= - \int_{a}^{b} f'(x) \phi(x) \; dx, \quad \text{as } \phi(a) = \phi(b) = 0.
\end{aligned}
\end{equation}
\textbf{Sobolev spaces} \cite{AdamsFournier2003SobolevSpaces}
\(W^{k, p}(\Omega)\) is a function space where all functions \(f\)
having weak derivatives up to order \(k\) and every derivate is
\(p\)-integrable via \eqref{eq:p-integrable},
\begin{equation}
W^{k, p}(\Omega) \subset L^p(\Omega) = \{ f: \Omega \subseteq \mathbb{R}^d \to \mathbb{R} \; | \; 
\forall \alpha \leq k. \; 
\exists \; \partial^{\alpha} f \in L^p(\Omega)  \},
\label{eq:sobolev-space}
\end{equation}
When \(k = 2\), it forms a Hilbert space,
i.e.~\(W^{k, 2}(\Omega) = H^k(\Omega)\).

\hypertarget{approximation-theory-of-neural-networks}{%
\subsection{Approximation theory of neural
networks}\label{approximation-theory-of-neural-networks}}

We quote some known bounds for neural networks from theoretical machine
learning field here, which are relevant to PDE solvers analysis later.

\textbf{Universal approximation theorem}
\cite{Hornik1989MLPUniversalApprox} states that neural networks
\(\hat{u}_{\theta}\), for which parameters \(\theta \in \Theta\), can
approximate any continuous functions \(u: \mathbb{R}^d \to \mathbb{R}\)
with little error \(\epsilon > 0\) in the \(p\)-norm of function space
\(\mathcal{U}\), with an extension to their differential operator
\(\mathcal{D}\),
\begin{equation}
\exists \; \theta \in \Theta.  \;
||\hat{u}_{\theta} - u||_{p} < \epsilon  \implies || \mathcal{D}[\hat{u}_{\theta}] - \mathcal{D}[u] ||_p  < \epsilon.
\label{eq:universal-approximation}
\end{equation}
\textbf{Optimal DNN functions approximation theorem}
\cite{Yarotsky2018OptimalReLU}. Assuming a continuous function
\(u \in W^{s, p}\) as defined in \eqref{eq:sobolev-space}, where
\(s \in \mathbb{N}^+\) is the smoothness of \(u\), there exists a neural
network \(\hat{u}_{\theta}\) with \(M\) parameters, such that the error
bound is,
\begin{equation}
|| \hat{u}_{\theta} - u||_{p} = O(M^{-\frac{s}{d}}),
\label{eq:optimal-approximation}
\end{equation}
where \(d\) is the input dimension of \(u\). It means intuitively that
the smoother and lower-dimensional the \(u\) is, the easier for it to be
approximated by a neural network \(\hat{u}_{\theta}\). For a fixed error
\(\epsilon\), the required number of parameters is
\(M = O(\epsilon^{-\frac{d}{s}})\), which suffers from the exponential
growth of \(d\), i.e.~\textbf{the curse of dimensionality}
\cite{Bellman1957DP}.

\hypertarget{error-analysis}{%
\subsection{Error analysis}\label{error-analysis}}

\textbf{Error and risk estimation}. Define the \emph{risk} of the
approximation \(\hat{u}_{\theta}\) against the ground truth function
\(u: \Omega \to \mathbb{R}\) with \(p\)-norm integral,
\begin{equation}
\mathcal{R}(\hat{u}_{\theta}) = (||\hat{u}_{\theta} - u||_{p})^p := \int_{\Omega} |\hat{u}_{\theta}(x) - u(x)|^p \; dx,
\label{eq:function-risk}
\end{equation}
For discretized computation, quadrature \(\hat{\mathcal{R}}\) is used to
approximate the integral risk \(\mathcal{R}\) with \(N\) sample points
from the dataset, where \(\omega_k\) is the weight at each sample index
\(k\),
\begin{equation}
\begin{aligned}
\underbrace{ \int_{\Omega} |\hat{u}_{\theta}(x) - u(x)|^p \; dx}_{\text{integral } \mathcal{R}(\hat{u}_{\theta})}
= & \underbrace{\sum_{k=1}^{N} \omega_{k} | \hat{u}_{\theta}(x_k) - u(x_k) |^p}_{\text{discretized } \hat{\mathcal{R}}(\hat{u}_{\theta})} + \mathcal{E}_{T}(\theta)
\approx
\hat{\mathcal{R}}(\hat{u}_{\theta}), \quad 
\mathcal{E}_{T}(\theta) := \mathcal{R}(\hat{u}_{\theta}) - \hat{\mathcal{R}}(\hat{u}_{\theta}),
\end{aligned}
\label{eq:training-error}
\end{equation}
where the training, also known as generalization or out-of-sample, error
\(\mathcal{E}_{T}(\theta)\) measures the difference between the integral
risk and the discretized risk due to quadrature.

\textbf{Error decomposition} \cite{kutyniok2022mathAI}. When
approximating a continuous function \(u\) with a neural network
\(\hat{u}_{\theta}\), the \emph{total error}
\(\mathcal{E}(\theta)\)\footnote{$\inf_\theta$ is the infimum over all possible network parameters $\theta$, which might not be attained.}
is decomposed into three parts, with the risk \(\mathcal{R}\) and its
quadrature \(\hat{\mathcal{R}}\) defined in \eqref{eq:function-risk} and
\eqref{eq:training-error} respectively,
\begin{equation}
\mathcal{E}(\theta):= \mathcal{R}(\hat{u}_{\theta})  
\leq
\underbrace{
   \inf_{\theta^{\star} \in \Theta} \mathcal{R}(u_{\theta^{\star}})
}_{\mathcal{E}_{A}(\theta)}
+ 
\underbrace{
    \hat{\mathcal{R}}(\hat{u}_{\theta}) - \inf_{\theta^{\star} \in \Theta} \mathcal{R}(u_{\theta^{\star}})
}_{\mathcal{E}_{O}(\theta)}
+
\underbrace{
    \mathcal{R}(\hat{u}_{\theta}) - \hat{\mathcal{R}}(\hat{u}_{\theta})
}_{\mathcal{E}_{T}(\theta)},
\label{eq:error-decomposition}
\end{equation}
the \emph{approximation error} \(\mathcal{E}_{A}\) measures the risk
between the best network approximation \(u_{\theta^{\star}}\) and ground
truth \(u\), \emph{optimization error} \(\mathcal{E}_{O}\) measures the
trained network result \(\hat{u}_{\theta}\) deviation from the best
network approximation, and \emph{training error} \(\mathcal{E}_{T}\)
defined in \eqref{eq:training-error}.

\hypertarget{pinn-learning-theory}{%
\subsection{PINN learning theory}\label{pinn-learning-theory}}

We briefly analyze the \acrshort{pinn} error
bound\footnote{The PDE residual is considered here, whereas boundary and initial conditions are omitted for simplicity.}.
The total error between the optimal solution \(\hat{u}_{\theta}\) and
the ground truth \(u\) is shown in \eqref{eq:pinn-total-error}. However,
during training, the network doesn't have access to the exact ground
truth for \(u\). Therefore, we aim to reduce the PDE residual instead.
\begin{equation}
\begin{aligned}
\mathcal{E}_{\mathcal{R}}(\theta) = (||\mathcal{R}_{\theta}||_p)^p 
&= (||\mathcal{D}[\hat{u}_{\theta}] - f||_p)^p, \quad \text{by } \eqref{eq:pinn-residual}. \\
&= || \mathcal{D}[\hat{u}_{\theta}] - \mathcal{D}[u] ||_p, \quad \text{by } \eqref{eq:pde-general}. \\
&= || \hat{f} - f ||_p, \quad \text{by the definition of } \hat{f},
\label{eq:pinn-residual-error}
\end{aligned}
\end{equation}
where \(\hat{f} = \mathcal{D}[\hat{u}_{\theta}]\) is the approximated
source function. In practice, this integral is approximated via
quadrature, with training error defined in
\eqref{eq:pinn-training-error}.

From the \textbf{theoretical} perspective, the goal is to derive that
the total error \(\mathcal{E}_{\text{PINN}}\)
\eqref{eq:pinn-total-error} is sufficiently small. To prove this, a
sufficient condition is that the total error is bounded by the residual
error \(\mathcal{E}_{\mathcal{R}}\) \eqref{eq:pinn-residual-error},
i.e.~we can prove that the smallest residual error ensures the smallest
total error.
\begin{equation}
\forall \theta \in \Theta. \; 
\mathcal{E}_{\text{PINN}}(\theta) \leq  C \mathcal{E}_{\mathcal{R}}(\theta),
\label{eq:error-bound-1}
\end{equation}
where \(C\) is a constant. By expansion of \(\mathcal{E}_{\text{PINN}}\)
\eqref{eq:pinn-total-error} and \(\mathcal{E}_{\mathcal{R}}\)
\eqref{eq:pinn-residual-error}, the abovementioned inequality
\eqref{eq:error-bound-1} is equivalent to the following
\textbf{coercivity} condition \cite{DeRyck2022ErrorAnalysisPINNs},
\begin{equation}
\forall \theta \in \Theta. \;  ||\hat{u}_{\theta} - u|| \leq C || \hat{f} - f ||_p,
\label{eq:pinn-coercivity}
\end{equation}
By quadrature bound \cite{Iserles2008NumericalAnalysis}, the smallest
practical training error \(\mathcal{E}_{T}\)
\eqref{eq:pinn-training-error} ensures the smallest residual error
\(\mathcal{E}_{R}\) \eqref{eq:pinn-residual-error}, where \(C'\) is a
constant,
\begin{equation}
\forall \theta \in \Theta. \; \mathcal{E}_{\mathcal{R}}(\theta) \leq C' [\mathcal{E}_{T}(\theta)+
\mathcal{E}_{u}(N_\mathcal{R})],
\label{eq:error-bound-2}
\end{equation}
and the extra term \(\mathcal{E}_{u}(N_\mathcal{R})\) converges faster
than \(\frac{1}{N_\mathcal{R}}\), thus can be ignored given the
increasing sampled quadrature points \(N_\mathcal{R}\),
\begin{equation}
\begin{aligned}
\mathcal{E}_{u}({N_\mathcal{R}}) \sim o(\frac{1}{{N_\mathcal{R}}}) & \implies \frac{\mathcal{E}_{u}({N_\mathcal{R}})}{\frac{1}{{N_\mathcal{R}}}} = 0, n \to \infty, & \quad \text{by the definition of little-o notation}. \\
& \implies \lim_{{N_\mathcal{R}} \to \infty} {N_\mathcal{R}} \mathcal{E}_{u}({N_\mathcal{R}}) = 0, & \quad \text{by the definition of limit}. 
\end{aligned}
\label{eq:little-o-analysis}
\end{equation}
By the above two inequalities \eqref{eq:error-bound-1} and
\eqref{eq:error-bound-2}, the total error \(\mathcal{E}_{\text{PINN}}\)
\eqref{eq:pinn-total-error} converges as the training error
\(\mathcal{E}_{T}\) \eqref{eq:pinn-training-error} converges,
\begin{equation}
\forall \theta \in \Theta. \;
\mathcal{E}_{\text{PINN}}(\theta) \leq C C' [\mathcal{E}_{T}(\theta) + o(\frac{1}{{N_\mathcal{R}}})]. 
\label{eq:total-error-bound}
\end{equation}
By Universal approximation theorem \eqref{eq:universal-approximation},
the smoothness of the solution \(u\) ensures that the residual error
\(\mathcal{E}_{\mathcal{R}}(\theta) < \epsilon\) is sufficiently small.
Given sufficient quadrature points \({N_\mathcal{R}}\), and smooth
activation functions in the neural network \(\hat{u}_{\theta}\)
\cite{Iserles2008NumericalAnalysis},
\begin{equation}
\min_{\theta \in \Theta} \mathcal{E}_{T}(\theta) \leq \mathcal{E}_{\mathcal{R}}(\theta) + o(\frac{1}{{N_\mathcal{R}}}),
\end{equation}
Hence, the training error
\(\mathcal{E}_{T}(\theta) < \epsilon + o(\frac{1}{{N_\mathcal{R}}})\) is
sufficiently small, according to \eqref{eq:little-o-analysis}. So is the
total error
\(\mathcal{E}_{\text{PINN}}(\theta) < C C' [\epsilon + o(\frac{1}{{N_\mathcal{R}}})]\),
by \eqref{eq:total-error-bound}, which concludes the proof.

From the \textbf{practical} perspective, the common failure modes, from
the above theoretical analysis, are (1) few quadrature points
\({N_\mathcal{R}}\) leading to large training error \(\mathcal{E}_{T}\)
in \eqref{eq:training-error}, (2) insufficient training resulting in
large optimization error \(\mathcal{E}_{O}\) in
\eqref{eq:error-decomposition}, (3) violation of the coercivity
condition \eqref{eq:pinn-coercivity} for PDEs, and (4) large constant
\(C\) in \eqref{eq:error-bound-1} or \(C'\) in \eqref{eq:error-bound-2}.

\end{document}